\documentclass[prl,twocolumn,superscriptaddress,showpacs,amsmath,amstex,amssymb,citeautoscript]{revtex4-1}
\pdfoutput=1
\usepackage{natbib}
\usepackage[english]{babel}
\usepackage{letltxmacro}
\usepackage{latexsym}
\LetLtxMacro{\ORIGselectlanguage}{\selectlanguage}
\makeatletter
\DeclareRobustCommand{\selectlanguage}[1]{%
  \@ifundefined{alias@\string#1}
    {\ORIGselectlanguage{#1}}
    {\begingroup\edef\x{\endgroup
       \noexpand\ORIGselectlanguage{\@nameuse{alias@#1}}}\x}%
}
\newcommand{\definelanguagealias}[2]{%
  \@namedef{alias@#1}{#2}%
}
\makeatother
\definelanguagealias{en}{english}
\definelanguagealias{English}{english}
\usepackage{graphicx}
\usepackage{amsmath}
\usepackage{amsfonts}
\usepackage{amssymb}
\usepackage{bm}
\usepackage{color}
\usepackage[percent]{overpic}
\usepackage{soul} 
\usepackage{amssymb}
\usepackage{wasysym}
\usepackage{dsfont}
\usepackage{float}
\usepackage[mathscr]{euscript}
\usepackage{lipsum}
\usepackage{hyperref}
\hypersetup{
    pdfstartview={FitH},    
    colorlinks=true,       
    linkcolor=blue,          
    citecolor=blue,        
    filecolor=magenta,      
    urlcolor=blue           
}

\usepackage{pdfpages}
\usepackage{pgffor}
\makeatletter
\AtBeginDocument{\let\LS@rot\@undefined}
\makeatother

\newcommand{\pdagger}{\phantom{\dagger}}
\newcommand{\be}{\begin{equation}}
\newcommand{\ee}{\end{equation}}
\newcommand{\bea}{\begin{eqnarray}}
\newcommand{\eea}{\end{eqnarray}}

\newcommand{\mc}{\mathcal}

\newcommand{\vect}[1]{\boldsymbol{#1}}

\usepackage{graphicx}
\usepackage[colorinlistoftodos]{todonotes}
\usepackage{verbatim}
\usepackage[normalem]{ulem}

\begin{document}

\title{Emergent $\mathbb{Z}_2$ gauge theories and topological excitations in Rydberg atom arrays}
\author{Rhine Samajdar}
\affiliation{Department of Physics, Harvard University, Cambridge, MA 02138, USA}
\author{Darshan G. Joshi}
\affiliation{Department of Physics, Harvard University, Cambridge, MA 02138, USA}
\author{Yanting Teng}
\affiliation{Department of Physics, Harvard University, Cambridge, MA 02138, USA}
\author{Subir Sachdev}
\affiliation{Department of Physics, Harvard University, Cambridge, MA 02138, USA}
\affiliation{School of Natural Sciences, Institute for Advanced Study, Princeton, NJ 08540, USA}

\begin{abstract}
Strongly interacting arrays of Rydberg atoms provide versatile platforms for exploring exotic many-body phases and dynamics of correlated quantum systems. Motivated by recent experimental advances, we show that the combination of Rydberg interactions and appropriate lattice geometries naturally leads to emergent $\mathbb{Z}_2$ gauge theories endowed with matter fields. Based on this mapping, we describe how Rydberg platforms could realize two distinct classes of topological $\mathbb{Z}_2$ quantum spin liquids, which differ in their patterns of translational symmetry fractionalization. We also discuss the natures of the fractionalized excitations of these $\mathbb{Z}_2$ spin liquid states using both fermionic and bosonic parton theories, and illustrate their rich interplay with proximate solid phases.

\end{abstract}
\maketitle

\hypersetup{linkcolor=blue}

\label{sec:rydberg}

Quantum spin liquids (QSLs) are strongly correlated phases of matter characterized by long-range many-body quantum entanglement, which gives rise to exotic properties such as fractionalized excitations, emergent gauge fields, and topological ground-state degeneracies \cite{savary2016quantum,knolle2019field,broholm2020quantum}. The simplest example of such a QSL which does not break any symmetries, including time-reversal, is the $\mathbb{Z}_2$ spin liquid  \cite{ReadSachdev91,Wen91,Sachdev92}---a stable, gapped quantum state with the same topological order as the toric code \cite{kitaev2006anyons}. Despite
some indications that such a phase may exist in certain electronic
systems on the kagome lattice \cite{lee2008end}, direct experimental detection thereof has so far proved elusive in solid-state materials. 

Recently, efforts towards realizing $\mathbb{Z}_2$ QSL phases have turned to programmable quantum simulators based on Rydberg atom arrays \cite{Browaeys.2020,morgado2021quantum,Semeghini.2021} which offer clean, tunable platforms to probe these highly entangled states and their dynamics.
Generically, the system consists of atoms individually trapped in an array of optical tweezers and pumped by lasers to highly excited Rydberg states.
The basic physics of this setup is that of the ``Rydberg blockade'' \cite{jaksch2000fast}: the interaction between two atoms in the Rydberg state is very large at short distances, and this significant energy cost prohibits---or \textit{blockades}---the simultaneous excitation of neighboring atoms, thereby inducing robust quantum correlations between the atomic states. This effect can be exploited to study a number of interesting phases of quantum matter and the transitions between them, thus prompting a wealth of experimental \cite{bernien2017probing,de2019observation,keesling2019quantum,Ebadi.2021,scholl2021quantum,bluvstein2021controlling} and theoretical \cite{samajdar2018numerical,whitsitt2018quantum,PhysRevLett.122.017205,chepiga2021kibble,Samajdar_2020,PhysRevLett.126.170603,kalinowski2021bulk,Kim.2021,orourke2022entanglement} investigation.

A promising playground to look for QSL phases is the family of quantum dimer models \cite{RK, moessner2008quantum}.
Recently, Ref.~\onlinecite{Samajdar.2021} showed that the phases of various quantum dimer models \cite{yan2022triangular} can be efficiently implemented using Rydberg atoms arrayed on the {\it sites\/} of a kagome lattice and argued that Rydberg platforms could be used to realize topological spin liquid states based solely on their native interactions. Recent experiments on an array of Rydberg atoms placed on the {\it links\/} of a kagome lattice yielded evidence for a state with topological correlations \cite{Semeghini.2021}, in accordance with theoretical proposals \cite{Verresen.2020}. However, even though numerics and experiments support a $\mathbb{Z}_2$ QSL phase, a general understanding of the connection betweeen Rydberg atom arrays and $\mathbb{Z}_2$ QSLs remains to be obtained.

In this article, we bridge the gap and establish the underlying reason \textit{why} geometrically frustrated Rydberg atom arrays host spin liquids. First, we show that the Rydberg Hamiltonian can be mapped to a $\mathbb{Z}_2$ gauge theory; the spin liquid phase is the deconfined phase of such a gauge theory. However, this emergent gauge theory is necessarily endowed with matter fields. These matter fields are the three distinct anyonic quasiparticle excitations of the $\mathbb{Z}_2$ QSL, which can be either bosonic ($e$ and $m$) or fermionic ($\varepsilon$). The $e$ and $\varepsilon$ anyons are particle-like excitations, and are collectively referred to as ``spinons'', whereas the $m$ anyon is a vortex-like excitation called a ``vison'' \cite{SenthilFisher}. Constructing detailed parton theories for each of these excitations, we analyze their static spectra using self-consistent mean-field theory and illustrate their relation to neighboring nontopological phases in the context of spinon condensation.

Importantly, depending on whether elementary translations anticommute or commute when acting on the visons, $\mathbb{Z}_2$ spin liquids can be further classified as ``odd'' or ``even'', respectively \cite{Jalabert91,sachdev2000translational,MSF02}. In
the $\mathbb{Z}_2$ gauge theory framework, there is a unit
$\mathbb{Z}_2$ electric charge on each lattice site of an odd $\mathbb{Z}_2$ gauge theory, which is a manifestation of nontrivial lattice symmetry fractionalization in this phase \cite{essin2013classifying,chen2015anomalous,tarantino2016symmetry,barkeshli2019symmetry}. The visons see the spinons as a source of $\pi$ flux, so the  adiabatic motion of a vison around a lattice site picks up a phase of $+1$\,($-1$) in an even (odd) QSL. We highlight how this subtle distinction is reflected in a parton formulation and adds to the rich variety of possible QSL states.

\begin{figure}[t]
\begin{center}
\includegraphics[width=\linewidth]{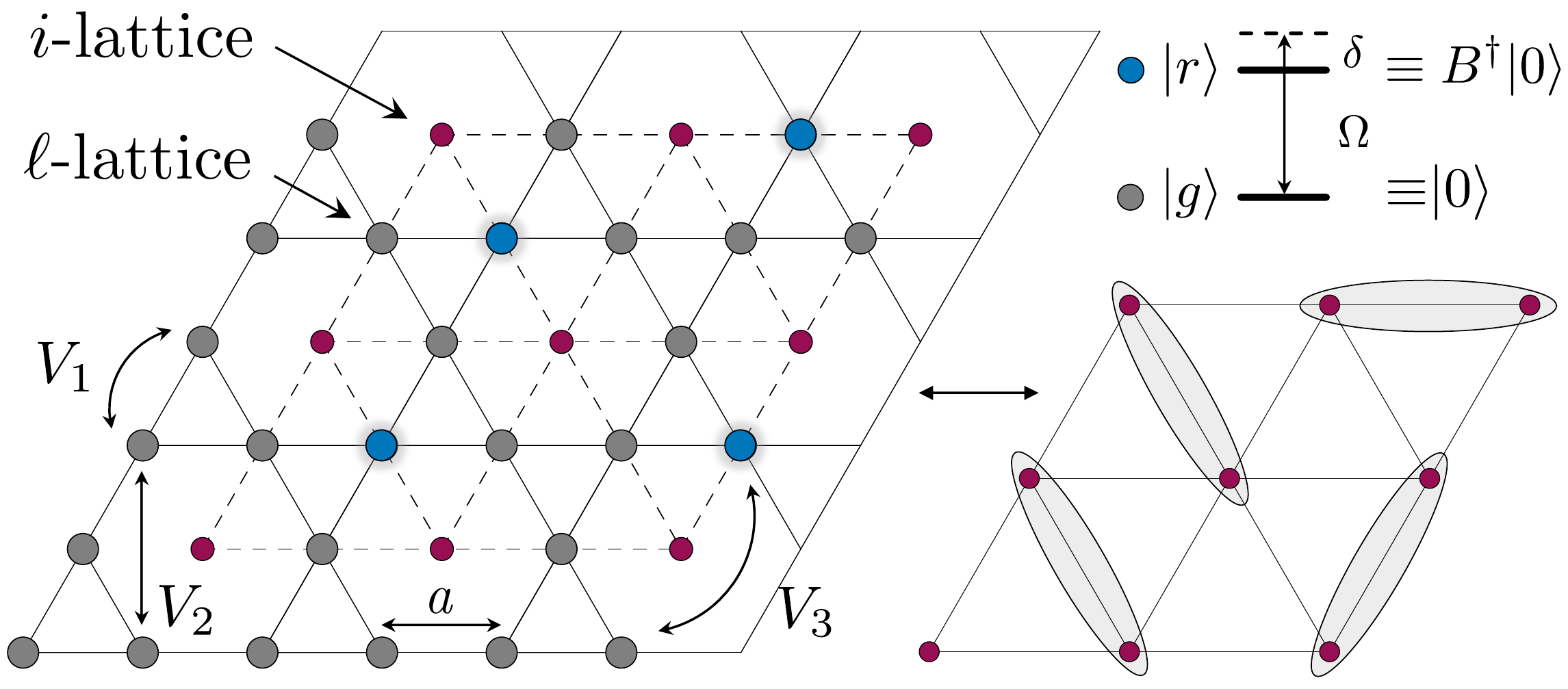}
\end{center}
\caption{Rydberg atoms are positioned on the sites of the $\ell$-lattice, which is a kagome lattice; the associated $i$-lattice (red dots) is triangular. Here, the two atomic states $\rvert g \rangle$ (gray) and $\rvert r \rangle$ (blue) are identified with an empty and occupied bosonic mode, respectively. The lattice spacing is $a$, and the interaction between $p$-th-nearest neighbors is represented by $V_p$. Any configuration of Rydberg excitations can be mapped to a set of dimers that need not satisfy a close-packing constraint.}
\label{fig:Rydberg_triangle}
\end{figure}

\textit{Model.---}In the simplest description, each Rydberg atom can be effectively regarded as a two-level system (i.e., a qubit).
We identify the atomic ground state $\left| g \right\rangle$ with an empty bosonic state $\left| 0 \right\rangle$ and the Rydberg state $\left| r \right\rangle$ with the filled bosonic state $B^\dagger \left| 0 \right\rangle$ (Fig.~\ref{fig:Rydberg_triangle}). By construction, this mapping associates the states with ``hard-core'' bosons, i.e.,
$
N^{\vphantom\dagger}_\ell \equiv B_\ell^\dagger B_\ell^{\vphantom\dagger} = 0, 1.
$

These two states are coupled by the external lasers with a Rabi frequency $\Omega$. The frequency of the laser is adjusted such that the detuning away from resonance of the $\left| g \right\rangle$ to $\left| r \right\rangle$ transition is $\delta$ (see~Fig.~\ref{fig:Rydberg_triangle}). Atoms in the Rydberg state interact via a van der Waals potential of the form $\mc{V} ({\bm r}) \equiv V_0/r^6$, which arises from strong dipole-dipole interactions. Putting these ingredients together, the full system can be described by a model originally proposed by Fendley, Sengupta, and Sachdev~(FSS) \cite{fendley2004competing,sachdev2002mott}. The Hamiltonian is given by
\begin{equation}
H^{}_{\textsc{fss}} = \sum_\ell \left[\frac{\Omega}{2}  \left( B_\ell^{\vphantom\dagger} + B_\ell^\dagger \right) - \delta N^{}_\ell \right] + \frac{1}{2} \sum_{\ell \neq \ell'}
V^{}_{\ell,\ell'} N^{}_\ell N^{}_{\ell'}, \label{HFSS}
\end{equation}
where $\ell$ labels a set points on the kagome lattice with position ${\bm r}_\ell$ and we have defined $V_{\ell, \ell'}$\,$\equiv$\,$\mc{V}(\vect{r}_\ell-\vect{r}_{\ell'})$ for notational brevity.
The interaction strength can equivalently be parametrized in terms of the Rydberg blockade radius  $R_b$\,$\equiv$\,$(V_0/\Omega)^{1/6}$; intuitively, this means that atoms within a radius of approximately $R_b$ are blockaded from occupying the Rydberg state simultaneously. The first term in the Hamiltonian \eqref{HFSS} breaks U($1$) symmetry, so the number of $B$ bosons is not conserved.

\textit{Emergent gauge theory.}---At the moment, $ H_{\textsc{fss}}$ is not a lattice gauge theory, and $B_{\ell}$ is the annihilation operator of a boson which does not carry gauge charges. We are interested here in configurations of the FSS model which can realize a $\mathbb{Z}_2$ spin liquid. To begin, we identify the two bosonic states on each site with the qubits of a $\mathbb{Z}_2$ gauge theory as
\bea
B^{\pdagger}_\ell + B_\ell^\dagger & = & \sigma^z_\ell,\quad
N^{\pdagger}_\ell   =  (1 - \sigma^x_\ell)/2. \label{BNsigma}
\eea
Then, without approximation, one can write the FSS model as a model of interacting qubits:
\begin{equation}
H^{}_{\textsc{fss}} = \frac{1}{2} \sum_\ell \left[ \Omega \, \sigma^z_\ell + \delta \, \sigma^x_\ell \right]  + \frac{1}{2} \sum_{\ell \neq \ell'}
\frac{V_{\ell,\ell'}}{4} (1 - \sigma^x_\ell)(1 - \sigma^x_{\ell'})\,. \label{HFSS2}
\end{equation}
In order to study possible $\mathbb{Z}_2$ spin liquid states, we explore making (\ref{HFSS2}) gauge invariant by introducing zero-energy matter fields which carry a $\mathbb{Z}_2$ gauge charge. First, we introduce an ``$i$-lattice'' of sites $i,j, \ldots$, such that the centers of the $(i,j)$ links on the $i$-lattice coincide with the $\ell$ sites in Eq.~(\ref{HFSS2}). Note that the $i$-lattice has to be defined in a manner which does not break any symmetries of the $\ell$-lattice. Such a construction is feasible for only some lattices---like the kagome \cite{Samajdar.2021} and the ruby \cite{Verresen.2020}---but not others; e.g., the square and honeycomb $\ell$-lattices do not have a corresponding $i$-lattice. 



Having found suitable $\ell$- and $i$-lattices, we place the Rydberg atoms on the $\ell$-lattice, and introduce a new set of qubits on the $i$-lattice, akin to the fractionalization schemes used for the pyrochlores \cite{PhysRevLett.108.037202, PhysRevB.86.104412}. 
The $i$-lattice qubits are acted on by Pauli matrices $\tau_i^{x,y,z}$, and these transform under $\mathbb{Z}_2$ lattice gauge transformations as
\begin{alignat}{3}
\sigma^z_{\bar{i}\bar{j}} &\rightarrow \varrho^{}_i  \sigma^z_{\bar{i}\bar{j}} \varrho^{}_j,\,\,\, \sigma^x_{\bar{i}\bar{j}} \rightarrow \sigma^x_{\bar{i}\bar{j}},\,\,\,
\tau^z_{i} \rightarrow \tau^z_{i} \varrho^{}_i ,\,\,\, \tau^x_{i} \rightarrow \tau^x_{i} \,.  \label{sigmataugauge2}
\end{alignat}
with $ \varrho_i$\,$=$\,$\pm 1$, where $\sigma^z_{\bar{i}\bar{j}} \equiv \sigma^z_{\ell}$ on the $\ell$-lattice site between the $i$ and $j$ sites on the $i$-lattice. 
Then, an explicitly $\mathbb{Z}_2$-gauge-invariant form of the FSS Hamiltonian is
\begin{equation}
\mathcal{H}^{}_{\textsc{fss}} = \frac{\Omega}{2} \sum_{\langle \bar{i}\bar{j} \rangle} \tau^z_i \sigma^z_{\bar{i}\bar{j}} \tau^z_j +\frac{\delta}{2} \sum_\ell  \sigma^x_\ell   + \sum_{\ell \neq \ell'}
\frac{V_{\ell, \ell'}}{8} (1 - \sigma^x_\ell)(1 - \sigma^x_{\ell'}). \label{HFSS3}
\end{equation}
The other canonical terms of Ising gauge theory, which are unessential to our discussion here, are described in Sec.~SI of the Supplemental Material (SM) \cite{supplement}.

With the introduction of the $\tau^z$ Ising matter fields, we also introduce an infinite number of gauge charges $G_i$ that commute with $\mathcal{H}_{\textsc{fss}}$ as
\begin{equation}
G^{}_i= \tau^x_i \prod_{\ell {\rm \,ends \, on\,} i} \sigma^x_{\ell} \,; \label{defGil}
\end{equation}
we choose $G_i$\,$=$\,$1$, whereupon the Hilbert space of Eq.~\eqref{HFSS3} is identical to that of Eq.~\eqref{HFSS}. Without dynamic matter, a state with $\tau^x_i$\,$=$\,$1$\,$(-1)$ will correspond to an even (odd) $\mathbb{Z}_2$ spin liquid; with dynamic matter, these identifications will continue to hold in a phase where $\tau^x_i$ has small fluctuations from the matter-free case.

\textit{Mean-field theory of bosonic spinons.---}Focusing hereafter on the case where the $\ell$-lattice is the kagome and the $i$-lattice is triangular (Fig.~\ref{fig:Rydberg_triangle}), we formulate a theory for the ground state and its $e$ excitations by returning to the bosonic description in Eq.~\eqref{BNsigma}. The $\tau_i^{x,z}$ operators can be similarly represented in terms of hard-core bosons $b$ such that
$
b_i^{\pdagger}$\,$+$\,$b_i^{\dagger}$\,$=$\,$\tau^z_i$, $b^\dagger_i b^{\pdagger}_i$\,$\equiv$\,$n^{\pdagger}_i$\,$=$\,$(1 \pm \tau_i^x)/2,
$
where the signs correspond to the odd/even cases, so that $\langle n_i$\,$\rangle$\,$\ll 1$ for both.
Then, the gauge charge operator in Eq.~\eqref{defGil} can be rewritten as
$
G^{}_i = \exp \left(i \pi \left[n^{}_i + \sum_{\ell {\rm \,ends \, on\,} i} N^{}_\ell \right] \right),
$
so we look for ground states with
\begin{equation}
n^{}_i + \sum_{\ell {\rm \,ends \, on\,} i} N^{}_\ell = 1,2.
\label{eq:constraint}
\end{equation}

\begin{figure}[H]
\centering
\includegraphics[width=\linewidth, trim={10cm 0cm 10cm 0cm},clip]{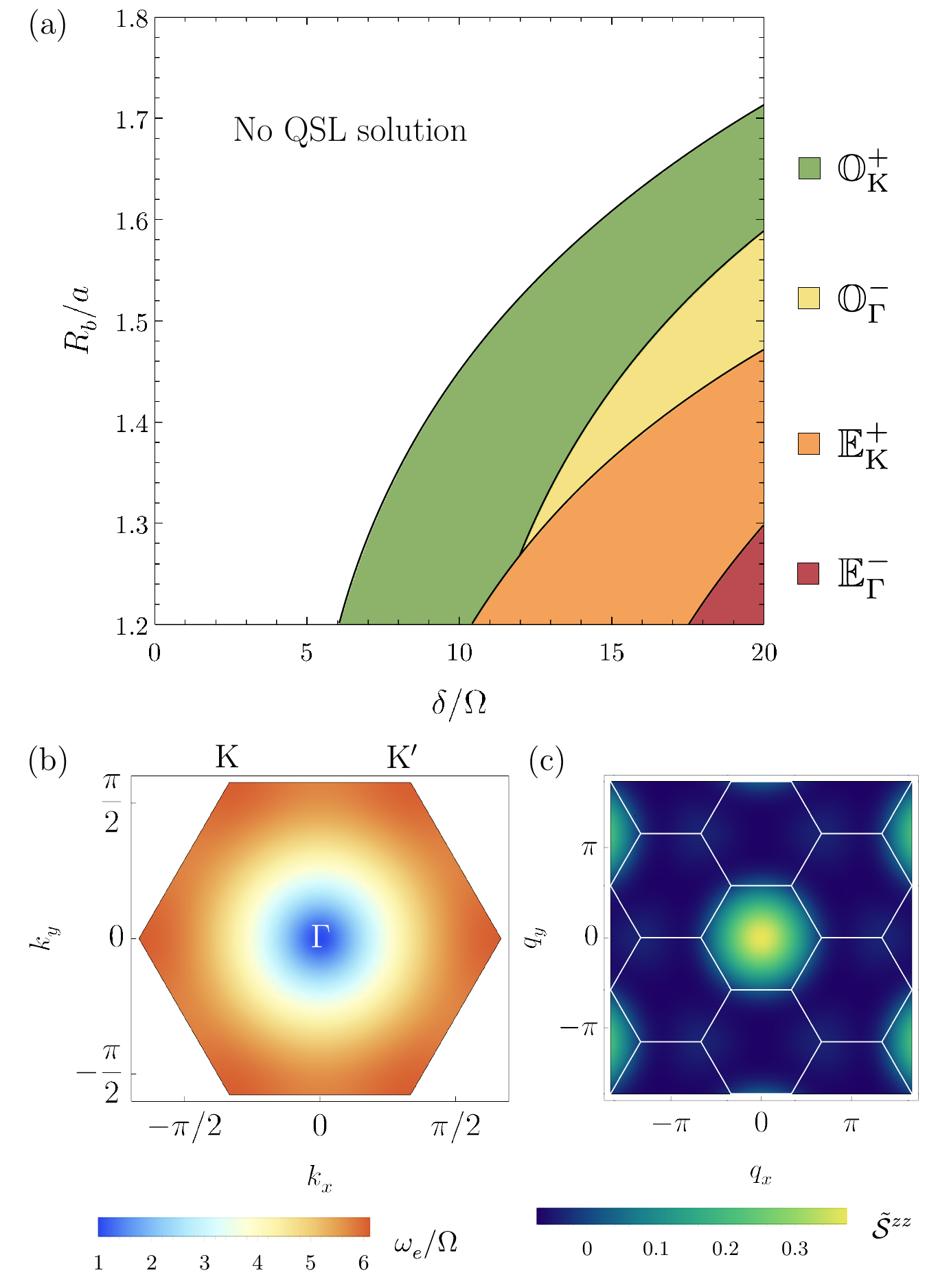}
    \caption{\label{fig:PD}(a) Mean-field phase diagram highlighting the four possible QSL solutions. (b) Representative energy dispersion of the $\mathbb{O}^{-}_{\Gamma}$ QSL in the first Brillouin zone. (c) The (approximate) static structure factor of this state displays prominent spectral weight at the $\Gamma$ point but is broadly featureless in the extended Brillouin zone (cf.~Fig.S2).}
\end{figure}

We can now perform a self-consistent mean-field theory calculation, after expressing Eq.~\eqref{HFSS3} in terms of $B_\ell$ and $b_i$,
and imposing the constraint in Eq.~\eqref{eq:constraint} by a Lagrange multiplier. In this process, which is described in detail in Sec.~SII \cite{supplement}, we condense $B_\ell$ and replace it with
a real number $\mc{B}$. Retaining terms quadratic in $b_i$ and diagonalizing the resulting Bogoliubov Hamiltonian, we arrive at the gapped $e$-particle spectrum.

The results of such an analysis are presented in Fig.~\ref{fig:PD} and Figs.~S1--S3 \cite{supplement}. In total, we find four distinct QSL states, which we label as $[\mathbb{E}/\mathbb{O}]^{[+/-]}_{[\Gamma/\mathrm{K}\,]}$. In this notation, $\mathbb{E}$ and $\mathbb{O}$ denote even and odd spin liquids, respectively, the $\pm$ in the superscript indicates the sign of $\mc{B}$ in the corresponding solution, and the subscript conveys whether the minima of the dispersion in the Brillouin zone occur at the $\Gamma$ point or at the $\mathrm{K}$, $\mathrm{K}'$ points. In Fig.~\ref{fig:PD}(a), we construct a mean-field phase diagram by plotting the lowest-energy solution among these four at each point in parameter space. While mean-field theory is not expected to capture the precise parameter values for QSL solutions, it does correctly describe the change in the nature of the QSL state from even to odd as the density of Rydberg excitations decreases with increasing $R_b/a$ \cite{yan2022triangular}. 

The representative spectra of the candidate Rydberg QSLs are shown in Fig.~\ref{fig:PD}(b) and Fig.~S2(a--c) \cite{supplement}. While all these states are gapped, one can reach an instability of the QSL state by tuning some parameter to bring the quasiparticle energy gap to zero. Then, the transition \textit{out} of the QSL into the proximate phases can be viewed as a condensation of the bosonic spinon \cite{Sachdev92,bais2009condensate}. For instance, consider the $\mathbb{O}^{-}_{\Gamma}$ QSL [Fig.~\ref{fig:PD}(b)]: since its dispersion minimum occurs at the $\Gamma$ point, when the $b_i$ are also condensed, one obtains a trivial paramagnetic or ``disordered'' phase as is commonly observed in the Rydberg phase diagram \cite{Samajdar.2021}. This quantum phase transition \cite{sachdev2011quantum} belongs to the so-called Ising$^*$ universality class \cite{chubukov1994universal,schuler2016universal,whitsitt2016transition}. Moreover, to investigate spin correlations in the QSL phase, in Figs.~\ref{fig:PD}(c) and S2, we analytically calculate an approximate static structure factor $\tilde{\mc{S}}^{zz} (\vect{q})$ in Fourier space based on the two-point functions $\langle S^z_i S^z_j\rangle$ for up to third-nearest-neighboring $i,j$ [Eq.~(S21)].
Since it only requires measurement of local observables, $\tilde{\mc{S}}^{zz} (\vect{q})$ provides a nontrivial experimentally accessible \cite{Semeghini.2021} prediction to probe and distinguish possible spin liquid states.

\begin{figure}[tb]
    \centering
    \includegraphics[width=\linewidth]{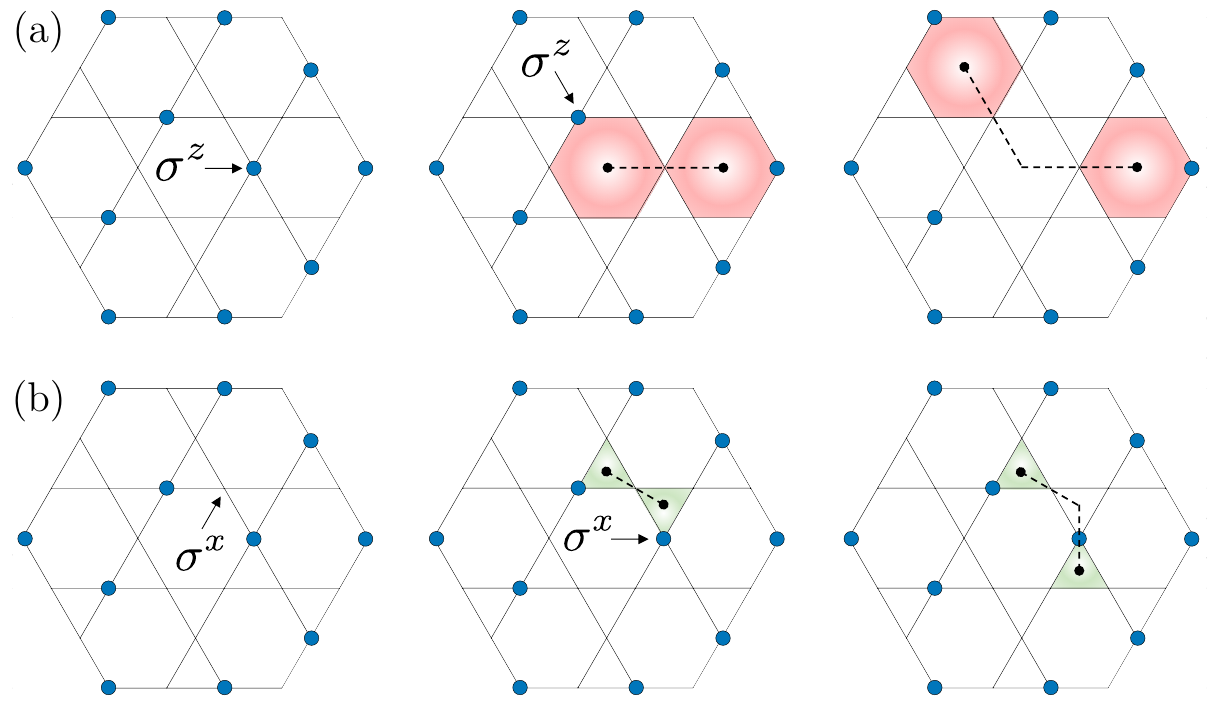}
    \caption{(a) Creation and motion of bosonic spinon ($e$ particle) excitations in an even QSL by the repeated application of $\sigma^z$, as depicted from left to right. The red plaquettes identify the defect hexagons on which the spinons reside. (b) Same, but for the visons ($m$ particles), which live at the centers of the kagome triangles. Both excitations can only be created in pairs by any local operator acting on the ground state. }
    \label{fig:KagomeEM}
\end{figure}

Pictorially, the Ising electric charge, which sits at the center of the hexagonal plaquettes of the kagome lattice, is defined by ``defect hexagons'' \cite{roychowdhury2015z} such that
$\prod_{\ell \,\in\,\hexagon} \sigma^x_\ell$\,$=$\,$-1$\,($+1$) for an even (odd) QSL,
as sketched in Fig.~\ref{fig:KagomeEM}(a). It is also easy to see from this figure why the gauge-charged matter fields $\tau^x$ are gapped.
Naively, given the presence of $\tau^z$-gapless matter, one could anticipate that $\tau^z$ would condense, destroying any possible $\mathbb{Z}_2$ QSL phase. However, from Fig.~\ref{fig:KagomeEM}(a), we notice that the motion of the Ising matter $\tau^x$ requires a $\sigma^z$ operation; by virtue of Eq.~\eqref{BNsigma}, this can add or remove a $B_\ell$ boson, leading to a large energetic cost from either $V({\bf r})$ or $\delta$, respectively. Consequently, $\tau^x$ gauge charge fluctuations are expensive, and this could help stabilize a deconfined phase of the $\mathbb{Z}_2$ gauge theory (\ref{HFSS3}).

\textit{Dual theory of visons.}---The second type of bosonic excitations of the $\mathbb{Z}_2$ QSL are the visons, which carry $\mathbb{Z}_2$ magnetic flux \cite{SenthilFisher,read1989statistics}. When there exists a conserved U($1$) charge, we conventionally regard the $e$ ($m$) particle as a boson with charge $Q$\,$=$\,$1/2$ ($Q$\,$=$\,$0$). However, in the absence of a conserved U(1) charge, as in our case, there is no sharp distinction between the $e$ and $m$ particles, but despite this nomenclatural ambiguity, the two are (more importantly) still relative semions.

For a full description of the $m$ particles, we perform a duality transformation on Eq.~\eqref{HFSS3} to obtain an Ising gauge theory with Ising matter on the lattice dual to the (triangular) $i$-lattice. This is the medial \textit{honeycomb} lattice formed by connecting the centers of the kagome triangles [Fig.~\ref{fig:KagomeEM}(b)]; on its sites, we define the Ising matter fields $\mu^z_{i'}$\,$=$\,$\pm1$, and on its links, we introduce the gauge fields $\eta^z_{i'j'}$\,$=$\,$\pm1$. The mapping between the direct $(\sigma, \tau)$ and dual $(\eta, \mu)$ variables is derived in Sec.~SIIIA \cite{supplement}, following which, we arrive at the theory for the visons:
\begin{alignat}{1}
\label{eq:vison}
&\tilde{\mc{H}}^{}_{\textsc{fss}} = \frac{\Omega}{2}\sum_{\langle \bar{i'}\bar{j'} \rangle} \eta^x_{\bar{i'}\bar{j'}} + \frac{\delta}{2} \sum_{\langle \bar{i'}\bar{j'} \rangle} \left(\mu^z_{i'} \eta^z_{\bar{i'}\bar{j'}}\mu^z_{j'} -1 \right)\\
\nonumber&+
\hspace*{-0.35cm}
\sum_{\langle \bar{i'}\bar{j'}\rangle \ne \langle \bar{k'}\bar{l'}\rangle} 
\hspace*{-0.35cm}
\frac{V(\vect{r}_{\bar{i'}\bar{j'}}^{}-\vect{r}_{\bar{k'}\bar{l'}}^{})}{8} \big(1 -  \mu^z_{i'} \eta^z_{\bar{i'} \bar{j'} }\mu^z_{j' } \big) \big(1 -  \mu^z_{k' } \eta^z_{\bar{k'}\bar{l'}}\mu^z_{l' } \big).
\end{alignat}

Upon restricting ourselves to only nearest-neighbor blockade interactions for simplicity and with the appropriate choice of a gauge (in the limit of low spinon densities), 
the minimal theory of the visons reduces to an Ising model on the honeycomb
lattice [Eq.~(S40)] with first- ($J_1$) and second-nearest-neighbor ($J_2$) Ising interactions (see Sec.~SIIIB \cite{supplement}). While not explicitly present in Eq.~\eqref{eq:vison}, we can also generically have a transverse-field term $\sim$\,$-K \sum_{i'} \mu^x_{i'}$ \cite{roychowdhury2015z} that arises in  perturbation theory [see Eq.~(S39)]. For large $K$, the $\mu$ spins
are polarized along the $x$ direction with a finite vison gap $\sim 2 K$. The underlying $\mu^z$ spins---and consequently, the boson number per site---are thus fluctuating, and this $\mu$-paramagnet
can be identified with the $\mathbb{Z}_2$ QSL \cite{roychowdhury2015z}. On the other hand, when $J_{1,2}$\,$\gg$\,$K$, the visons acquire
nontrivial dispersion \cite{moessner2001ising}, and if the minima thereof touches zero,
they can condense leading to $\langle \mu^z_i \rangle \ne0$ and the onset of long-range order \cite{huh2010, huh2011vison}. We emphasize, however, that any observable order parameter has to be composed of a \textit{pair} of visons.

For the even QSL, \citet{roychowdhury2015z} demonstrated that in the presence of a third-nearest-neighbor interaction $J_3$ (which would arise from the long-ranged Rydberg tails in our case), there is an extended regime in $J_{1,2,3}$ parameter space where the minima of the vison spectra occur at the three inequivalent $\mathrm{M}$ points in the Brillouin zone. Pairwise condensation of these visons would then describe the  transition to a threefold-rotational-symmetry-breaking ``nematic'' phase of Rydberg atoms on the kagome lattice \cite{Samajdar.2021,roychowdhury2015z}, characterized by ordering wavevectors at $2$M$_{1,2,3}$. Furthermore, in Sec.~SIIIB \cite{supplement}, we show that for the odd QSL, the minima of the vison dispersion are also consistent with the development of nematic order by spinon condensation, but can additionally reproduce a subset of the ordering peaks of a proximate ``staggered'' phase \cite{Samajdar.2021,roychowdhury2015z}.

\textit{Fermionic spinons.}---The anyon content of the $\mathbb{Z}_2$ QSL also includes a fermionic spinon. 
In order to obtain a theory of this $\varepsilon$ particle, we use the Abrikosov fermion representation \cite{abrikosov1965electron, affleck19882, marston1989large}, in which the spin operator at each site is fractionalized as
$
\vec S^{}_\ell$\,$\equiv$\,$\vec \sigma^{}_\ell/2$\,$=$\,$(f_\ell^\dagger\,\vec\rho \,f_\ell^{\pdagger})/2,
$
with $f_\ell$\,$\equiv$\,$(f_{\ell,1}, f_{\ell,2})^\mathrm{T}$ being a two-component fermionic spinon operator, and $\vec\rho$ denoting the three Pauli matrices. 

\begin{figure}[t]
    \centering
    \includegraphics[width=\linewidth]{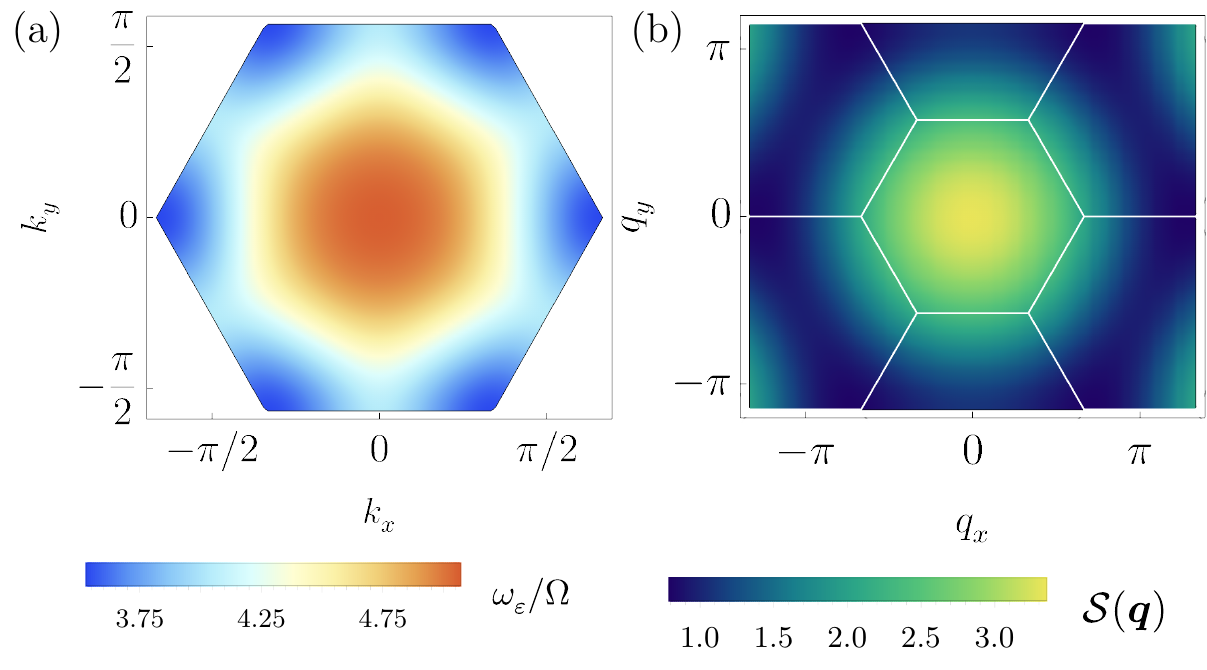}
    \caption{(a) Mean-field spectrum of the $\varepsilon$ particle and (b) the resultant static structure factor at $\delta/\Omega$\,$=$\,$4.0$, $R_b/a$\,$=$\,$1.60$. The qualitative nature of the fermionic band structure is the same for all ($\delta/\Omega$, $R_b/a$) for which a QSL solution is found.}
    \label{fig:epsilon}
\end{figure}

Writing $H^{}_{\textsc{fss}}$ in Eq.~\eqref{HFSS} in terms of these fermionic spinons generates four-fermion terms, which we decouple into fermion bilinears by introducing the mean-field parameters $t_{\ell, \ell'}^{\alpha\beta} $\,$\equiv$\,$\langle f_{\ell,\alpha}^\dagger f_{\ell',\beta}^{\pdagger}\rangle$, $\Delta_{\ell, \ell'}^{\alpha\beta}$\,$\equiv$\,$ \langle f_{\ell,\alpha}^{\pdagger} f_{\ell',\beta}^{\pdagger}\rangle$. The expectation values $\{t_{\ell, \ell'}^{\alpha\beta}, \Delta_{\ell, \ell'}^{\alpha\beta} \}$ then collectively
define a mean-field \textit{ansatz}. The projective action \cite{wen2002quantum, essin2013classifying}  of
lattice or time-reversal symmetries on this ansatz describes
the particular QSL state of interest. Unlike systems with SU($2$) spin-rotation symmetry (for which $\langle f^\dagger_{i,\alpha}f^{\pdagger}_{j,\beta}\rangle \propto \delta_{\alpha \beta}$ and $\langle f_{i,\alpha}f_{j,\beta}\rangle \propto \epsilon_{\alpha \beta}$ \cite{lu2011z}), in our ansatz, we have to allow for hopping and pairing terms with all possible combinations of $\alpha$,\,$\beta$ due to the lack of spin-rotation invariance in Eq.~\eqref{HFSS2}. The full theory thus obtained is detailed in Sec.~SIV of the SM \cite{supplement} [see Eq.~(S65)].
Self-consistently solving for $\{t_{\ell, \ell'}^{\alpha\beta}, \Delta_{\ell, \ell'}^{\alpha\beta} \}$ then yields the fermionic band structure. As illustrated in Fig.~\ref{fig:epsilon}(a), we observe that the $\varepsilon$ particle is gapped too and that the minima of its dispersion occur at the $\mathrm{K},\mathrm{K}'$ points. To determine the experimental signatures of this QSL state, we also calculate its static structure factor in Fig.~\ref{fig:epsilon}(b) and find that it has broad features located at $\vect{q}$\,$=$\,$\Gamma$ but no sharp Bragg peaks anywhere in the extended Brillouin zone, indicating the absence of long-range order.

\textit{Discussion and outlook.}---In this work, we have shown how systems of Rydberg atoms arrayed on kagome or ruby lattices can give rise to emergent $\mathbb{Z}_2$ gauge theories with matter fields; the deconfined phase of such a gauge theory is a $\mathbb{Z}_2$ quantum spin liquid. We develop a formalism to systematically characterize all three classes of  topological excitations of this $\mathbb{Z}_2$ QSL, evaluate their spectral properties in a parton description [Eqs.~\eqref{HFSS3}, \eqref{eq:vison}, and (S65)], and discuss their experimental fingerprints in static structure factors. In particular, we identify a promising QSL candidate, labeled $\mathbb{O}_\Gamma^-$, which is consistent with observations of neighboring nematic and disordered phases in the framework of spinon/vison condensation \cite{Samajdar.2021}. Our results bear direct relevance to recent and ongoing experiments on Rydberg quantum simulators that have opened the door to realizing and probing these highly entangled $\mathbb{Z}_2$ QSL states. Even if the exact ground state of the Rydberg system happens to not be a QSL, one can \textit{dynamically} prepare such a state in experiments via quasiadiabatic sweeps \cite{Semeghini.2021}, and our analysis herein of  the quasiparticle spectra and gaps should also help inform the feasibility of such processes.  

\begin{acknowledgments}
We thank Tom Banks, Yuan-Ming Lu, Mikhail Lukin, Zi Yang Meng, Sergej Moroz, and our coauthors in prior collaborations \cite{Samajdar.2021,Semeghini.2021,yan2022triangular} for useful discussions. This work was supported by the U.S. Department of Energy under Grant DE-SC0019030.  D.G.J. acknowledges support from the Leopoldina fellowship by the German National Academy of Sciences through Grant No.~LPDS 2020-01.
\end{acknowledgments}

\bibliographystyle{apsrev4-2_custom}
\bibliography{RydIGT_Refs}

\newpage


\section*{Supplementary material}

# Supplemental Material for "Emergent $\mathbb{Z}_2$ gauge theories and topological excitations in Rydberg atom arrays"

Rhine Samajdar,[1] Darshan G. Joshi,[1] Yanting Teng,[1] and Subir Sachdev[1, 2]

[1]*Department of Physics, Harvard University, Cambridge, MA 02138, USA*
[2]*School of Natural Sciences, Institute for Advanced Study, Princeton NJ 08540, USA*

## SI.  ISING GAUGE THEORY HAMILTONIAN

In this section, we systematically derive the different canonical terms of an Ising gauge theory with matter fields [1] in the context of the Rydberg system. To briefly review our discussion in the main text, we have seen that an explicitly $\mathbb{Z}_2$-gauge-invariant form of the FSS Hamiltonian describing the Rydberg atom array is given by

$$\mathcal{H}_{\textsc{fss}} = \frac{\Omega}{2} \sum_{\langle ij \rangle} \tau_i^z \sigma_{ij}^z \tau_j^z + \frac{\delta}{2} \sum_{\ell} (\sigma_\ell^x - 1) \quad \text{(S1)}$$
$$+ \frac{1}{2} \sum_{\ell_1 \neq \ell_2} \frac{V(\boldsymbol{r}_{\ell_1} - \boldsymbol{r}_{\ell_2})}{4} (1 - \sigma_{\ell_1}^x)(1 - \sigma_{\ell_2}^x).$$

Here, the relevant qubits lie on two distinct lattices, with the "$\ell$-lattice" hosting the $\sigma$ (gauge) fields and the "$i$-lattice" the $\tau$ (matter) fields. As before, we use the notation $\sigma_{ij}^z$ to denote $\sigma_\ell^z$ on the $\ell$-lattice site between the $i$ and $j$ sites on the $i$-lattice. Upon introducing matter fields, we now have an infinite number of gauge charges $G_i$ that commute with $\mathcal{H}_{\textsc{fss}}$ as

$$G_i = \tau_i^x \prod_{\ell \text{ ends on } i} \sigma_\ell^x. \quad \text{(S2)}$$

We pick a set of charges such that $G_i = 1 \, \forall \, i$; in other words, we define $\tau_i^x = \pm 1$ as the Ising matter vacuum, and $\tau_i^x = \mp 1$ as the "spinon" excitation for the even/odd gauge theory, respectively. Then, to enforce the condition that the ground state has no Ising matter, we can add a gauge-invariant term to the Hamiltonian

$$\mathcal{H}_\tau = \mp \gamma \sum_i \tau_i^x, \quad \text{(S3)}$$

with $\gamma > 0$, and study the theory in the limit $\gamma \to 0$.

It is also useful to recall that we have already argued for $\tau^x$ gauge charge fluctuations being expensive [Fig. 3(a)], which could help stabilize a deconfined phase of the $\mathbb{Z}_2$ gauge theory (S1). Consequently, we can eliminate the $\tau^z$ matter fields in an expansion in $\Omega$, and this will induce terms involving the gauge-invariant product of $\sigma_\ell^z$ around closed loops:

$$\mathcal{H}_{\text{loop}} = -\sum_{\text{loops}} K_{\text{loop}} \prod_{\ell_1, \ell_2, \ell_3 \ldots \in \text{loop}} \sigma_{\ell_1}^z \sigma_{\ell_2}^z \sigma_{\ell_3}^z \ldots \quad . \quad \text{(S4)}$$

Such terms suppress fluctuations in $\mathbb{Z}_2$ flux, and thus, stabilize a deconfined phase. Interestingly, this term also has an important interpretation in a picture where each Rydberg excitation (i.e., $\sigma_\ell^x = +1$) is identified with a *dimer* on the triangular lattice (Fig. 1) [2]; in this language, the $\mathbb{Z}_2$-flux terms of Eq. (S4) are then the dimer-flipping terms of the resultant quantum dimer model.

Putting together these individual ingredients, the sum $\mathcal{H}_{\textsc{fss}} + \mathcal{H}_\tau + \mathcal{H}_{\text{loop}}$ in Eqs. (S1), (S3), and (S4) contains all the canonical terms of $\mathbb{Z}_2$ gauge theory with Ising matter, along with the additional interaction terms $\propto V_{\ell,\ell'}$.

## SII.  BOSONIC SPINON EXCITATIONS

Using the framework of self-consistent mean-field theory, in this section, we compute the spectrum of the bosonic spinon, i.e., the $e$ particle, and classify the different possible QSL states. Our starting point is the gauge-invariant form of the FSS model in Eq. (S1). For concreteness, we focus here on the case where the $\ell$-lattice is a kagome lattice, on which the actual Rydberg atoms are positioned, whereas the $i$-lattice is triangular and hosts the $e$ spinons. However, we note that for a kagome $\ell$-lattice, it is also consistent to choose a honeycomb $i$-lattice. It remains an interesting and open question to see whether the theory for the $e$ particles on the honeycomb lattice could lead to any new states beyond those obtained here for the triangular lattice.

With the correspondence between spin and boson operators introduced in the main text

$$B_\ell + B_\ell^\dagger = \sigma_\ell^z, \quad B_\ell^\dagger B_\ell = (1 - \sigma_\ell^x)/2, \quad \text{(S5)}$$
$$b_i + b_i^\dagger = \tau_i^z, \quad b_i^\dagger b_i = (1 \pm \tau_i^x)/2, \quad \text{(S6)}$$

the model of Eq. (5) can be written, without approximation as

$$\mathcal{H}_{\textsc{fss}} = \frac{\Omega}{2} \sum_{\substack{\ell \\ i,j \,\in\, \partial \ell}} \left( b_i + b_i^\dagger \right) \left( B_\ell + B_\ell^\dagger \right) \left( b_j + b_j^\dagger \right) \quad \text{(S7)}$$
$$+ \frac{1}{2} \sum_{\ell_1 \neq \ell_2} V\left( \boldsymbol{r}_{\ell_1} - \boldsymbol{r}_{\ell_2} \right) \left( B_{\ell_1}^\dagger B_{\ell_1} \right) \left( B_{\ell_2}^\dagger B_{\ell_2} \right)$$
$$- \delta \sum_\ell B_\ell^\dagger B_\ell + \lambda \sum_i \left( b_i^\dagger b_i + \sum_{\ell \,|\, i \in \partial \ell} B_\ell^\dagger B_\ell - \mathcal{Q} \right).$$

In the last line, we have introduced a Lagrange multiplier $\lambda$ (with $\mathcal{Q} = 1, 2$) to enforce the constraint on the gauge charges in Eq. (7) of the main text, and the notation $i \in \partial \ell$ conveys that the link $\ell$ terminates on the

site $i$. To proceed with our mean-field description, we first condense $B_\ell$, and replace it with a real number $\mathcal{B} = \langle B \rangle = \langle B^\dagger \rangle$. This results in the Hamiltonian

$$\mathcal{H}_e = \Omega \mathcal{B} \sum_{\langle i,j \rangle} \left( b_i b_j + b_i^\dagger b_j^\dagger + b_i^\dagger b_j + b_i b_j^\dagger \right) + \sum_i \lambda b_i^\dagger b_i + \mathcal{C}(\lambda), \tag{S8}$$

with the constant $\mathcal{C}$ given by

$$\mathcal{C}(\lambda) = \mathcal{N}_\ell \mathcal{B}^4 (2V_1 + 2V_2 + 3V_3) - \delta \mathcal{N}_\ell \mathcal{B}^2 + \lambda \mathcal{N}_s \left( 6\mathcal{B}^2 - \mathcal{Q} \right),$$

where $\mathcal{N}_\ell$ and $\mathcal{N}_s$ are the number of sites on the kagome and triangular lattices, respectively ($\mathcal{N}_\ell = 3 \mathcal{N}_s$). The coefficients $V_p$ represent the strengths of the $p^{\text{th}}$-nearest-neighbor van der Waals interactions, which we truncate beyond $p = 3$. At this point, it is convenient to switch to Fourier space as

$$b_i = \frac{1}{\sqrt{\mathcal{N}_s}} \sum_{\boldsymbol{k}} e^{-i\boldsymbol{k} \cdot \boldsymbol{r}_i} b_{\boldsymbol{k}}. \tag{S9}$$

The triangular lattice has a lattice constant of $2a$, where $a$ is the spacing between adjacent Rydberg atoms on the kagome lattice; thus, $\boldsymbol{k}$ lies in a hexagonal Brillouin zone defined by the reciprocal lattice vectors $\boldsymbol{b}_1 = (1, -1/\sqrt{3})(\pi/a)$ and $\boldsymbol{b}_2 = (0, 2/\sqrt{3})(\pi/a)$. Henceforth, we will work in units where $a = 1$ unless specified otherwise. Substituting Eq. (S9) in Eq. (S8), the momentum-space Hamiltonian reads

$$\mathcal{H}_e = \frac{\Omega \mathcal{B}}{2} \sum_{\boldsymbol{k}} \left[ \gamma(\boldsymbol{k}) b_{\boldsymbol{k}} b_{-\boldsymbol{k}} + \gamma(\boldsymbol{k}) b_{\boldsymbol{k}}^\dagger b_{-\boldsymbol{k}}^\dagger + 2\gamma(\boldsymbol{k}) b_{\boldsymbol{k}}^\dagger b_{\boldsymbol{k}} \right]$$
$$+ \lambda \sum_{\boldsymbol{k}} b_{\boldsymbol{k}}^\dagger b_{\boldsymbol{k}} + \mathcal{C}(\lambda), \tag{S10}$$

where

$$\gamma(\boldsymbol{k}) = 2 \left( 2 \cos(k_x) \cos\left(\sqrt{3} k_y\right) + \cos(2k_x) \right) = \gamma(-\boldsymbol{k}) \tag{S11}$$

is the structure factor of the triangular lattice. This Hamiltonian can now be straightforwardly diagonalized by a Bogoliubov transformation. Introducing new creation and annihilation operators according to the relation

$$b_{\boldsymbol{k}}^\dagger = u_{\boldsymbol{k}} \beta_{\boldsymbol{k}}^\dagger + v_{-\boldsymbol{k}} \beta_{-\boldsymbol{k}}; \quad [\beta_{\boldsymbol{k}}, \beta_{\boldsymbol{k}'}^\dagger] = \delta_{\boldsymbol{k}, \boldsymbol{k}'}, \tag{S12}$$

for $u_{\boldsymbol{k}} = \cosh(\alpha_{\boldsymbol{k}})$, $v_{-\boldsymbol{k}} = \sinh(\alpha_{\boldsymbol{k}})$, and choosing $\alpha_{\boldsymbol{k}}$ judiciously, the Hamiltonian reduces to

$$\mathcal{H}_e = \sum_{\boldsymbol{k}} \omega(\boldsymbol{k}) \left( \beta_{\boldsymbol{k}}^\dagger \beta_{\boldsymbol{k}} + \frac{1}{2} \right) + \lambda \mathcal{N}_s \left( 6\mathcal{B}^2 - \mathcal{Q} - \frac{1}{2} \right)$$
$$+ \mathcal{N}_\ell \mathcal{B}^4 (2V_1 + 2V_2 + 3V_3) - \delta \mathcal{N}_\ell \mathcal{B}^2 \tag{S13}$$

with the dispersion relation $\omega_e(\boldsymbol{k}) = \sqrt{\lambda [\lambda + 2\Omega \mathcal{B} \gamma(\boldsymbol{k})]}$.

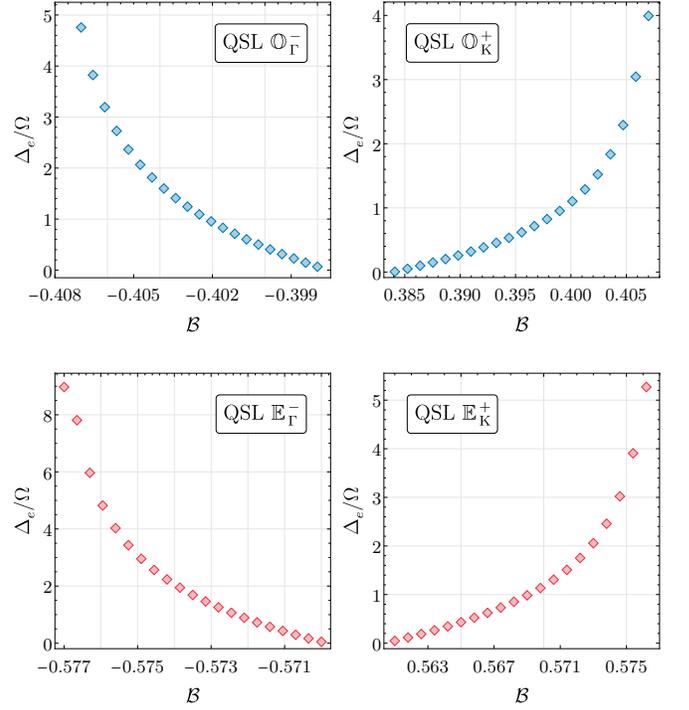

Figure S1. The energy gap as a function of the boson condensate $\mathcal{B}$ for the four possible QSL solutions. As $\mathcal{B}$ is varied, one can transition out of a given QSL ground state when either the $b$ boson condenses (i.e., the gap vanishes) or the self-consistency conditions in Eqs. (S14, S17) break down.

### A. Self-consistent spin-liquid solutions

Let us first look at the case with $\mathcal{Q} = 1$, which corresponds to the odd $\mathbb{Z}_2$ spin liquid. At zero temperature, minimizing $E_0 = \langle \mathcal{H}_e \rangle$ with respect to $\lambda$, we obtain the self-consistency condition

$$\frac{1}{3\mathcal{N}_s} \sum_{\boldsymbol{k}} \frac{\lambda + \Omega \mathcal{B} \gamma(\boldsymbol{k})}{\sqrt{\lambda^2 + 2\lambda \Omega \mathcal{B} \gamma(\boldsymbol{k})}} = 1 - 4\mathcal{B}^2. \tag{S14}$$

As $\lambda \to \pm \infty$, we see that the left-hand side of this equation tends to $\pm 1/3$ from above/below; this caps the possible values of $1 - 4\mathcal{B}^2$ for which a solution can exist as

$$\mathcal{B}^2 < \frac{1}{6} \text{ if } \lambda > 0, \quad \text{or} \quad \mathcal{B}^2 > \frac{1}{3} \text{ if } \lambda < 0. \tag{S15}$$

However, for $|\mathcal{B}| > 1/\sqrt{3}$, the constraint in Eq. (7), $n_i + 6\mathcal{B}^2 = \mathcal{Q}$, cannot be satisfied for positive $n_i$, so any such solutions should be discarded. At the same time, we must have $\lambda^2 + 2\lambda \Omega \mathcal{B} \gamma(\boldsymbol{k}) > 0 \; \forall \; \boldsymbol{k}$ in order for the left-hand side of Eq. (S14) to be real, which imposes the conditions

$$\begin{aligned} \lambda < -12\mathcal{B} \quad \text{or} \quad \lambda > 6\mathcal{B} & \quad \text{if} \quad \mathcal{B} > 0, \\ \lambda < 6\mathcal{B} \quad \text{or} \quad \lambda > -12\mathcal{B} & \quad \text{if} \quad \mathcal{B} < 0, \end{aligned} \tag{S16}$$

bounding the regions in $(\mathcal{B}, \lambda)$-parameter space where a self-consistent solution may be found. On combining the

| QSL | Constraint on $\mathcal{B}$ | Constraint on $\lambda$ |
|---|---|---|
| $\mathbb{O}_\Gamma^-$ | $-\frac{1}{\sqrt{6}} < \mathcal{B} < 0$ | $\lambda < -12\mathcal{B}$ |
| $\mathbb{O}_K^+$ | $0 < \mathcal{B} < +\frac{1}{\sqrt{6}}$ | $\lambda > 6\mathcal{B}$ |
| $\mathbb{E}_\Gamma^-$ | $-\frac{1}{\sqrt{3}} < \mathcal{B} < 0$ | $\lambda < -12\mathcal{B}$ |
| $\mathbb{E}_K^+$ | $0 < \mathcal{B} < +\frac{1}{\sqrt{3}}$ | $\lambda > 6\mathcal{B}$ |

Table I. Summary of the four self-consistent QSL solutions, and their regimes of existence in $(\mathcal{B}, \lambda)$-parameter space.

equations (S15) and (S16), we arrive at the two QSL solutions listed in Table I. Note that the Hamiltonian (S13) is not invariant under $\mathcal{B} \to -\mathcal{B}$, so solutions with different signs of $\mathcal{B}$ are indeed distinct.

Similarly, in the case of the even $\mathbb{Z}_2$ spin liquid for which $\mathcal{Q} = 2$, we find the consistency condition

$$\frac{1}{\mathcal{N}_s} \sum_{\boldsymbol{k}} \frac{\lambda + \Omega \mathcal{B} \gamma(\boldsymbol{k})}{\sqrt{\lambda^2 + 2\lambda \Omega \mathcal{B} \gamma(\boldsymbol{k})}} = 5 - 12\mathcal{B}^2. \quad (S17)$$

As before, we will have one set of constraints arising from the upper or lower bounds of the left-hand side of this equation:

$$\mathcal{B}^2 < \frac{1}{3} \text{ if } \lambda > 0, \quad \text{or} \quad \mathcal{B}^2 > \frac{1}{2} \text{ if } \lambda < 0, \quad (S18)$$

but solutions with $|\mathcal{B}| > 1/\sqrt{2}$ violate the constraint (7) and are unphysical; the requirements stemming from the reality of the energy denominator remain unchanged from Eq. (S16). This defines a further two even QSL solutions (see Table I), thus bringing the total to four.

For each of these QSL states, we compute the energy gap $\Delta_e$ and the full dispersion of the lowest bosonic band, which are illustrated in Figs. S1 and S2, respectively. As noted in the main text, the minima of the dispersion always occur at either the $\Gamma$ point or the $K, K'$ points, and this information is conveyed in our labeling of the different states.

### B. Correlation functions

A common observable to fingerprint different types of QSL states is the static structure factor. One can define such a quantity from the Fourier-transformed two-point correlation functions as

$$\mathcal{S}^{zz}(\boldsymbol{q}) = \frac{1}{\mathcal{N}_\ell} \sum_{\ell_1, \ell_2} e^{i\boldsymbol{q} \cdot (\boldsymbol{r}_{\ell_1} - \boldsymbol{r}_{\ell_2})} \langle \sigma_{\ell_1}^z \sigma_{\ell_2}^z \rangle, \quad (S19)$$

where, for the Rydberg system, the expectation value is computed in the ground state of the FSS Hamiltonian.

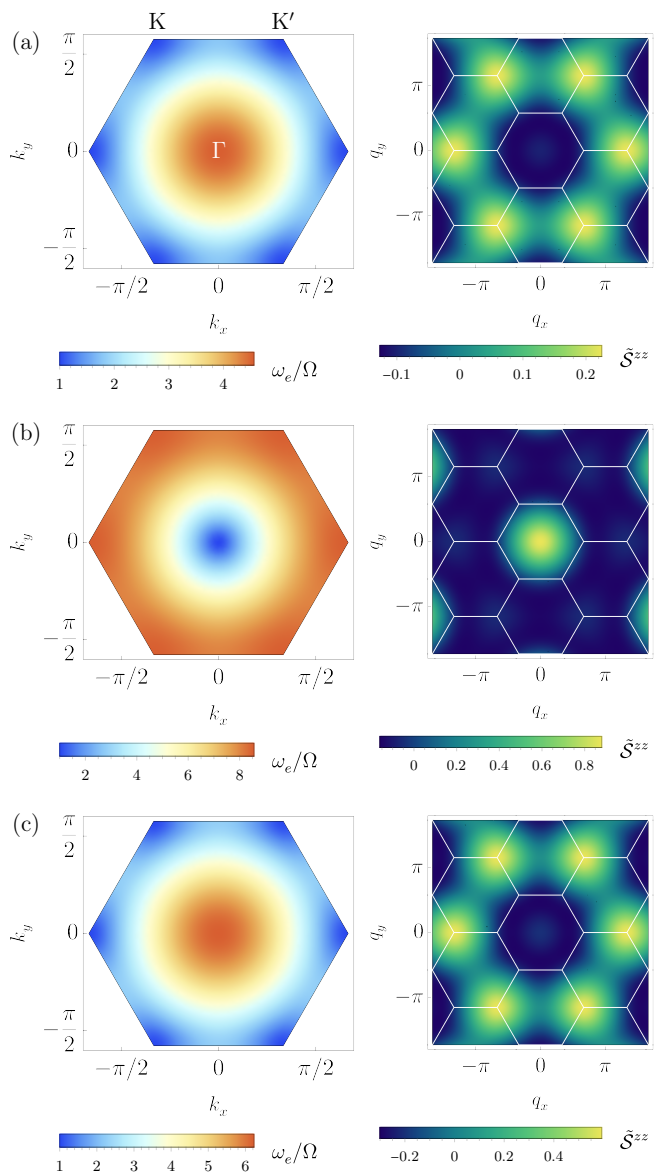

Figure S2. Left: Energies of the lowest bosonic band in the first Brillouin zone for the three QSLs (a) $\mathbb{O}_K^+$, (b) $\mathbb{E}_\Gamma^-$, and (c) $\mathbb{E}_K^+$. Right: the approximate structure factors $\tilde{\mathcal{S}}^{zz}(\boldsymbol{q})$ plotted over an extended region in reciprocal space; the white hexagons mark the dimensions of the first Brillouin zone. In each case, the value of $\mathcal{B}$ is chosen such that the gap equals one (in units of $\Omega$).

Identifying the action of the Rydberg $\sigma^z$ operator on the gauge and matter fields from Fig. 3(a), we see that in the gauge-theory language, the structure factor (S19) can be evaluated as

$$\mathcal{S}^{zz}(\boldsymbol{q}) = \frac{1}{2\mathcal{N}_\ell} \sum_{\substack{\langle i,j \rangle \\ \langle c,d \rangle}} e^{i\boldsymbol{q} \cdot (\boldsymbol{r}_{ij} - \boldsymbol{r}_{cd})} \langle \tau_i^z \sigma_{ij}^z \tau_j^z \tau_c^z \sigma_{cd}^z \tau_d^z \rangle, \quad (S20)$$

wherein $\langle i, j \rangle$ and $\langle c, d \rangle$ denote nearest-neighbor pairs on the triangular lattice, and the expectation value is now

calculated with respect to the Bogoliubov vacuum; note that the operator enclosed within $\langle \cdots \rangle$ is gauge invariant. When expressed in terms of the $b$ bosons, in general, this expression involves 16 four-boson terms and is challenging to evaluate. However, we can define an *approximate* structure factor by restricting the sum in Eq. (S19) to first-, second-, and third-nearest-neighbor correlators, i.e.,

$$\tilde{\mathcal{S}}^{zz}(\boldsymbol{q}) \equiv \frac{1}{2\mathcal{N}_\ell} \sum_{\substack{\langle i,j\rangle, \langle c,d\rangle \\ ||\boldsymbol{r}_{ij}-\boldsymbol{r}_{cd}||\leq 2a}} e^{i\boldsymbol{q}\cdot(\boldsymbol{r}_{ij}-\boldsymbol{r}_{cd})} \langle \tau_i^z \sigma_{ij}^z \tau_j^z \tau_c^z \sigma_{cd}^z \tau_d^z \rangle. \tag{S21}$$

Then, Eq. (S20) simplifies to the form

$$\tilde{\mathcal{S}}^{zz} = \frac{(2\mathcal{B})^2}{\mathcal{N}_\ell} \left[ 2 \sum_{\langle i,j\rangle} e^{i\boldsymbol{q}\cdot(\boldsymbol{r}_i-\boldsymbol{r}_j)/2} \langle \tau_i^z \tau_j^z \rangle \right. \tag{S22}$$

$$\left. + 2 \sum_{\langle\langle i,j\rangle\rangle} e^{i\boldsymbol{q}\cdot(\boldsymbol{r}_i-\boldsymbol{r}_j)/2} \langle \tau_i^z \tau_j^z \rangle + \sum_{\langle\langle\langle i,j\rangle\rangle\rangle} e^{i\boldsymbol{q}\cdot(\boldsymbol{r}_i-\boldsymbol{r}_j)/2} \langle \tau_i^z \tau_j^z \rangle \right],$$

which can be evaluated exactly in the bosonic formulation to obtain

$$\tilde{\mathcal{S}}^{zz} = \frac{4\mathcal{B}^2}{3\mathcal{N}_s} \sum_{\boldsymbol{k}} \left(|v_{\boldsymbol{k}}|^2 + u_{\boldsymbol{k}} v_{\boldsymbol{k}}\right) \left[ 2 \sum_{\boldsymbol{\Delta}_1} e^{-i\boldsymbol{q}\cdot\boldsymbol{\Delta}_1/2} \cos(\boldsymbol{k}\cdot\boldsymbol{\Delta}_1) \right.$$

$$\left. + 2 \sum_{\boldsymbol{\Delta}_2} e^{-i\boldsymbol{q}\cdot\boldsymbol{\Delta}_2/2} \cos(\boldsymbol{k}\cdot\boldsymbol{\Delta}_2) + \sum_{\boldsymbol{\Delta}_3} e^{-i\boldsymbol{q}\cdot\boldsymbol{\Delta}_3/2} \cos(\boldsymbol{k}\cdot\boldsymbol{\Delta}_3) \right], \tag{S23}$$

with the summation on $\boldsymbol{\Delta}_p$ running over all $p^{\text{th}}$-nearest-neighbor lattice vectors. The quantity $\tilde{\mathcal{S}}^{zz}(\boldsymbol{q})$ thus encodes information about the equal-time correlations and can be useful as an experimental probe to distinguish the various QSL phases. Specifically, in Figs. 2(c) and S2, we notice that the approximate structure factor exhibits broad features located at the $K, K'$ points of the *second* Brillouin zone for the $\mathbb{E}/\mathbb{O}_K^+$ QSLs, whereas for the $\mathbb{E}/\mathbb{O}_\Gamma^-$ spin liquids, the spectral weight is mostly concentrated around the $\Gamma$ point in the first Brillouin zone.

### C. Mean-field phase diagram

So far, we have regarded $\mathcal{B}$ as a free parameter. However, in reality, the value of $\mathcal{B}$ is set by some combination of the external parameters, $\delta/\Omega$ and $R_b/a$, and for each such point in parameter space, there exists an optimal $\mathcal{B}$. On minimizing the ground-state energy $E_0$ with respect to $\mathcal{B}$, we find the condition

$$\frac{1}{12\mathcal{N}_s} \sum_{\boldsymbol{k}} \frac{\lambda \Omega \gamma(\boldsymbol{k})}{\sqrt{\lambda^2 + 2\lambda \Omega \mathcal{B} \gamma(\boldsymbol{k})}} = \mathcal{B} \left(\delta - 2\lambda - 2\mathcal{V}\mathcal{B}^2\right),$$

using the shorthand $\mathcal{V} = 2V_1 + 2V_2 + 3V_3$. For every value of $\delta$ and $R_b$ (which fixes the interaction strengths $V_p$), this equation is to be solved for a self-consistent value of $\mathcal{B}$, which in turn determines $\lambda(\mathcal{B})$ according to Eq. (S14) or (S17). Depending on the resultant combination of $\mathcal{B}$ and $\lambda(\mathcal{B})$, there may exist one, multiple, or no QSL solutions; putting together this information allows one to construct a phase diagram in $(\delta/\Omega, R_b/a)$-space, as shown in Fig. 2(a). Of course, we also have to account for the possibility that the true ground state for a given $(\delta/\Omega, R_b/a)$ may not be a QSL (depending on the energetics of the competing phases), which lies beyond the scope of such a phase diagram.

## SIII. VISON EXCITATIONS

### A. Duality transformation

Before diving into the mapping between Eqs. (5) and (8) of the main text, it is useful to first recall the standard Wegner duality in $d=2$ relating the square-lattice quantum Ising model and pure lattice gauge theory *without* matter fields [1, 3, 4]. The familiar transverse-field Ising model (TFIM) is defined by the Hamiltonian

$$H_{\text{TFIM}} = -J \sum_{\langle i,j\rangle} \mu_i^z \mu_j^z - \gamma \sum_j \mu_j^x, \tag{S24}$$

where $\langle i,j\rangle$ denotes nearest neighbors, and $\mu$ are Ising spins residing on the sites of a square lattice (blue circles in Fig. S3), which we will hereafter refer to as the direct lattice. The Pauli operators $\mu_i^z$ ($\mu_i^x$) measure (flip) the state of the qubit at site $i$. Next, we introduce the dual lattice located at the centers of plaquettes of the direct lattice as shown by the red sites in Fig. S3. In this case, the dual lattice is also a square lattice, and we label its sites by a primed index, e.g., $s'$. The links of the direct and dual lattices are in one-to-one correspondence, so without ambiguity, we can use the notation $\mathcal{O}_{\bar{i}\bar{j}}$ ($\mathcal{O}_{\bar{i}'\bar{j}'}$) to denote an operator $\mathcal{O}$ living on the link connecting sites $i$ ($i'$) and $j$ ($j'$) of the direct (dual) lattice.

In order to proceed with our duality mapping, we introduce the dual variables $\sigma$ that live on the links of the dual lattice and are defined as

$$\sigma_{\bar{i}\bar{j}}^x = \mu_i^z \mu_j^z, \quad \mu_i^x = \sigma_{\bar{i}\bar{g}}^z \sigma_{\bar{i}\bar{h}}^z \sigma_{\bar{i}\bar{j}}^z \sigma_{\bar{i}\bar{l}}^z. \tag{S25}$$

Thus, $\sigma^x$ measures whether there is a domain wall between two neighboring spins on the direct lattice, while the second relation shows that flipping a single spin is equivalent to creating four domain walls along a closed loop. Using this correspondence, the TFIM in Eq. (S24) is dual to the Ising gauge theory

$$H_{\text{IGT}} = -J \sum_\ell \sigma_\ell^x - \gamma \sum_{\Box_i} \prod_{\ell \in \Box_i} \sigma_\ell^z, \tag{S26}$$

where the sum on $\ell$ runs over all links, and we have used the shorthand $\Box_i$ to denote a plaquette (of the dual

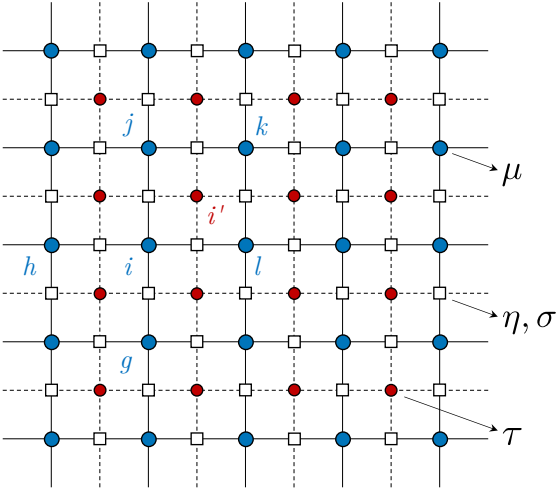

Figure S3. The direct square lattice (solid lines) and its dual (dashed lines), which is also a square lattice. The qubits $\mu$ and $\eta$ live on the sites (blue circles) and links (white squares) of the direct lattice, respectively, whereas the variables for the dual theory, $\tau$ and $\sigma$, reside on the sites (red circles) and links (white squares) of the dual lattice. Note that the links of the direct and dual lattices coincide.

lattice) centered on the site $i$; for instance, in Fig. S3, $\prod_{\ell \in \Box_i} \sigma_\ell^z \equiv \sigma_{\bar{i}\bar{g}}^z \sigma_{\bar{i}h}^z \sigma_{\bar{i}j}^z \sigma_{\bar{i}l}^z$. This theory is to be supplemented with the local constraint

$$\prod_{\ell \in +_{s'}} \sigma_\ell^x = 1 \ \forall \text{ dual lattice sites } s', \tag{S27}$$

where $+_{s'}$ denotes the set of links that emerge from the dual lattice site $s'$; for example, using Eq. (S25), it is easy to see that $\prod_{\ell \in +_{i'}} \sigma_\ell^x \equiv \sigma_{\bar{i}j}^x \sigma_{\bar{j}k}^x \sigma_{\bar{k}l}^x \sigma_{\bar{l}i}^x = 1$, by definition, for the sites labeled in Fig. S3. Since there are two links for every site of the direct lattice, this constraint is necessary to recover the correct number of degrees of freedom. Note that Eq. (S26) is invariant under the $\mathbb{Z}_2$ gauge transformation

$$\sigma_{\bar{i}'\bar{j}'}^z \to \varrho_{i'} \, \sigma_{\bar{i}'\bar{j}'}^z \, \varrho_{j'}, \quad \sigma_{\bar{i}'\bar{j}'}^x \to \sigma_{\bar{i}'\bar{j}'}^x, \tag{S28}$$

with $\varrho_{i'} = \pm 1$ being the elements of the gauge group.

Now, let us consider a system with two sets of Pauli matrices on links $\ell$, $\sigma_\ell^a$, and on sites $s'$, $\tau_{s'}^a$ with the Hamiltonian

$$\begin{aligned}\mathcal{H} = &-J \sum_\ell \sigma_\ell^x - \kappa \sum_p \prod_{\ell \in \Box_p} \sigma_\ell^z \\ &- \bar{\kappa} \sum_{s'} \tau_{s'}^x - K \sum_\ell \sigma_\ell^z \prod_{s' \in \partial\ell} \tau_{s'}^z,\end{aligned} \tag{S29}$$

together with the constraint that

$$\tau_{s'}^x \prod_{\ell \in +_{s'}} \sigma_\ell^x = 1 \ \forall \text{ dual lattice sites } s'. \tag{S30}$$

Keeping with our earlier notation, we will regard these qubits to be residing on the dual lattice and we want to find a mapping that gives us the theory defined on the sites and links of the direct square lattice. Notice that here, we are starting from a gauge theory equipped with a gauge constraint (S30), and applying the Wegner duality discussed above in "reverse". Introducing qubits $\mu$ and $\eta$ which live on the sites and links of the direct lattice, respectively, such a mapping is defined by

$$\begin{aligned}\sigma_{\bar{i}j}^x &= \mu_i^z \, \eta_{\bar{i}j}^z \, \mu_j^z, \quad \mu_i^x = \sigma_{\bar{i}\bar{g}}^z \sigma_{\bar{i}h}^z \sigma_{\bar{i}j}^z \sigma_{\bar{i}l}^z = \prod_{\ell \in \Box_i} \sigma_\ell^z, \\ \eta_{\bar{i}'\bar{j}'}^x &= \tau_{i'}^z \, \sigma_{\bar{i}'\bar{j}'}^z \, \tau_{j'}^z, \quad \tau_{i'}^x = \eta_{\bar{i}j}^z \eta_{\bar{j}k}^z \eta_{\bar{k}l}^z \eta_{\bar{l}i}^z = \prod_{\ell \in +_{i'}} \eta_\ell^z.\end{aligned} \tag{S31}$$

For the new variables, $\eta^z$ and $\mu^z$ probe the qubits while $\eta^x$ and $\mu^x$ flip them. Under this transformation, the Hamiltonian (S29) reads

$$\begin{aligned}\tilde{\mathcal{H}} = &-J \sum_\ell \eta_\ell^z \prod_{i \in \partial\ell} \mu_i^z - \kappa \sum_i \mu_i^x \\ &- \bar{\kappa} \sum_{p'} \prod_{\ell \in \Box_{p'}} \eta_\ell^z - K \sum_\ell \eta_\ell^x,\end{aligned} \tag{S32}$$

i.e., the $(2+1)$D Ising gauge theory with matter fields is self-dual [5, 6]. Note that the constraint (S30) has already been absorbed into the definition of $\tau^x$ in Eq. (S31) because

$$\tau_{i'}^x \prod_{\ell \in +_{i'}} \sigma_\ell^x = \left(\mu_i^z \mu_j^z \mu_k^z \mu_l^z\right)^2 \prod_{\ell \in +_{i'}} \eta_\ell^z \prod_{\ell \in +_{i'}} \eta_\ell^z = 1,$$

so it is identically satisfied in the dual theory. This is in analogy to the Wegner duality without matter fields, in which $H_{\text{IGT}}$ was accompanied by a constraint but its counterpart, $H_{\text{TFIM}}$, was not.

With this intuition from the square lattice, we can now construct the duality mapping between theories on the triangular and honeycomb lattices. Consider a set of qubits $\{\sigma_\ell^a, \tau_i^a\}$ ($\{\eta_{\ell'}^a, \mu_{i'}^a\}$) which are situated on links and sites of the triangular (honeycomb) lattice, respectively. As before, we will use the convention that $\sigma$ and $\eta$ represent the gauge fields whereas $\tau$ and $\mu$ stand for the matter fields. Then, in exact analogy to Eq. (S31), we define

$$\begin{aligned}\sigma_{\bar{i}'\bar{j}'}^x &= \mu_{i'}^z \, \eta_{\bar{i}'\bar{j}'}^z \, \mu_{j'}^z, \quad \mu_{i'}^x = \prod_{\ell \in \triangle_{i'}} \sigma_\ell^z, \\ \eta_{\bar{i}j}^x &= \tau_i^z \, \sigma_{\bar{i}j}^z \, \tau_j^z, \quad \tau_i^x = \pm \prod_{\ell' \in \bigcirc_i} \eta_{\ell'}^z.\end{aligned} \tag{S33}$$

Here, $\triangle_{i'}$ represents the triangular-lattice plaquette centered on the site $i'$ while $\bigcirc_i$ denotes the honeycomb lattice's hexagon centered on the site $i$. The $\pm$ in the last equation arises from the combination of Eq. (S2) with the (easily verifiable) relation

$$\prod_{\ell \in \bigcirc_i} \sigma_\ell^x = \prod_{\ell' \in \bigcirc_i} \eta_{\ell'}^z, \tag{S34}$$

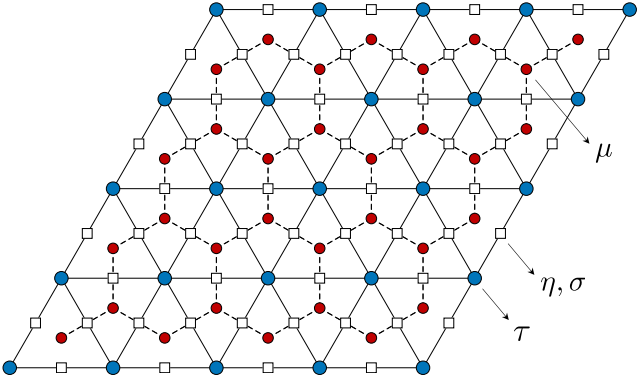

Figure S4. The direct triangular lattice (solid lines) and its dual honeycomb lattice (dashed lines). The qubits $\tau$ and $\sigma$ live on the sites (blue circles) and links (white squares) of the direct lattice, respectively, whereas the variables for the dual theory, $\mu$ and $\eta$, reside on the sites (red circles) and links (white squares) of the dual lattice. Note that the links of the direct and dual lattices once again coincide.

so the minus sign holds for the odd QSL. With the mapping (S33) applied to Eq. (S1), we find that the vison excitations of the $\mathbb{Z}_2$ spin liquid are described by the theory

$$\tilde{\mathcal{H}}_{\text{FSS}} = \frac{\Omega}{2} \sum_{\langle \bar{i}'\bar{j}' \rangle} \eta^x_{\bar{i}'\bar{j}'} + \frac{\delta}{2} \sum_{\langle \bar{i}'\bar{j}' \rangle} \left( \mu^z_{i'} \eta^z_{\bar{i}'\bar{j}'} \mu^z_{j'} - 1 \right) \quad (S35)$$
$$+ \sum_{\langle \bar{i}'\bar{j}' \rangle \neq \langle \bar{k}'\bar{l}' \rangle} \frac{V(\bm{r}_{\bar{i}'\bar{j}'} - \bm{r}_{\bar{k}'\bar{l}'})}{8} \bigl(1 - \mu^z_{i'}\eta^z_{\bar{i}'\bar{j}'}\mu^z_{j'}\bigr)\bigl(1 - \mu^z_{k'}\eta^z_{\bar{k}'\bar{l}'}\mu^z_{l'}\bigr).$$

Since the electric charges are gapped, at low energies, there are no spinons at all for the even Ising gauge theory, while there is a background of exactly one spinon per site in the odd $\mathbb{Z}_2$ gauge theory. By virtue of Eq. (S34), this requires that

$$\prod_{\ell' \in \bigcirc_i} \eta^z_{\ell'} = \begin{cases} +1 & \text{for an even } \mathbb{Z}_2 \text{ QSL,} \\ -1 & \text{for an odd } \mathbb{Z}_2 \text{ QSL,} \end{cases} \quad (S36)$$

where the product, as earlier, is over all six links of the hexagonal plaquette centered at site $i$.

### B. Vison dispersion and condensation

As seen from the last term in Eq. (S35), the theory $\tilde{\mathcal{H}}_{\text{FSS}}$ involves interactions between six qubits. In order to make analytical progress, let us focus on only nearest-neighbor interactions, which is a reasonable approximation since the dominant energy scale of the problem is set by the Rydberg blockade. This brings Eq. (S35) to the form (up to constants)

$$\tilde{\mathcal{H}}_0 = \frac{\Omega}{2} \sum_{\langle \bar{i}'\bar{j}' \rangle} \eta^x_{\bar{i}'\bar{j}'} + \left(\frac{\delta}{2} - V_1\right) \sum_{\langle i',j' \rangle} \mu^z_{i'} \eta^z_{\bar{i}'\bar{j}'} \mu^z_{j'} \quad (S37)$$

$$+ \frac{V_1}{4} \sum_{\langle\langle i',j' \rangle\rangle} \mu^z_{i'} \eta^z_{\bar{i}'\bar{k}'} \eta^z_{\bar{k}'\bar{j}'} \mu^z_{j'},$$

where $k'$ is the intermediate honeycomb vertex on the shortest path connecting second-nearest neighbors $i'$ and $j'$. On the second line, the additional factor of 2 multiplying $V_1$ vis-à-vis Eq. (S35) is because $\langle\langle i',j' \rangle\rangle$ represents the sum over all *unordered* next-nearest-neighbor pairs (and likewise for $\langle i',j' \rangle$).

As it currently stands, $\tilde{\mathcal{H}}_0$ is bereft of terms containing $\mu^x$, which generates fluctuations of $\mu^z$. However, such a term does arise from $\mathcal{H}_{\text{loop}}$ in Eq. (S4). Physically, $\mathcal{H}_{\text{loop}}$ describes a process where a spinon located at some site on the triangular lattice is taken around a loop by the repeated application of $\sigma^z$ [see Fig. 3(a)] and brought back to its starting point. Importantly, since the hopping parameter is parametrically small in $\Omega/\delta$, $K_{\text{loop}}$ falls off quickly for larger loops. Hence, the leading contribution in the sum (S4) comes from the shortest possible loop—this is constituted by the three spins forming a triangle on the kagome lattice. Thus, to a good approximation,

$$\mathcal{H}_{\text{loop}} \approx -K_\triangle \sum_{i'} \prod_{\ell_1, \ell_2, \ell_3 \in \triangle_{i'}} \sigma^z_{\ell_1} \sigma^z_{\ell_2} \sigma^z_{\ell_3}, \quad (S38)$$

which, by Eq. (S33), is dual to

$$\tilde{\mathcal{H}}_{\text{field}} = -K_\triangle \sum_{i'} \mu^x_{i'}. \quad (S39)$$

Taken together, $\tilde{\mathcal{H}}_{\text{FSS}} + \tilde{\mathcal{H}}_{\text{field}}$ describes the full theory for the visons.

#### 1. Even $\mathbb{Z}_2$ spin liquid

Owing to the no-spinon constraint in Eq. (S36), we can now choose a gauge where $\eta^z_{\ell'} = +1 \; \forall \; \ell'$. With this choice, the Hamiltonian for the visons is simply an Ising model with first- and second-nearest-neighbor couplings $(J_{1,2})$ and a transverse-field term given by Eq. (S39) [7]. When $J_{1,2} \gg K_\triangle$, the visons gain dispersion, and for sufficiently large $J_{1,2}$, there may be an ordering transition associated with the macroscopic occupation of the softest modes [8]. Neglecting constants, to zeroth order in $\mathcal{O}(K_\triangle/J_{1,2})$, Eq. (S37) reduces to

$$\tilde{\mathcal{H}}_0^{\text{even}} = J_1 \sum_{\langle i',j' \rangle} \mu^z_{i'} \mu^z_{j'} + J_2 \sum_{\langle\langle i',j' \rangle\rangle} \mu^z_{i'} \mu^z_{j'}, \quad (S40)$$

where $J_1 \equiv \delta/2 - V_1$, $J_2 \equiv V_1/4$. At large blockade radii, the competing ferromagnetic $(J_1 > 0)$ and antiferromagnetic $(J_2 < 0)$ couplings induce frustration in this classical model [9].

In order to determine the soft modes corresponding to the vison condensation transition, let us work in momentum space with

$$\mu^z_{i,\alpha} = \frac{1}{\sqrt{2\mathcal{N}_s}} \sum_{\bm{k}} e^{-i\bm{k}\cdot\bm{r}_i} \mu^z_{\bm{k},\alpha}, \quad (S41)$$

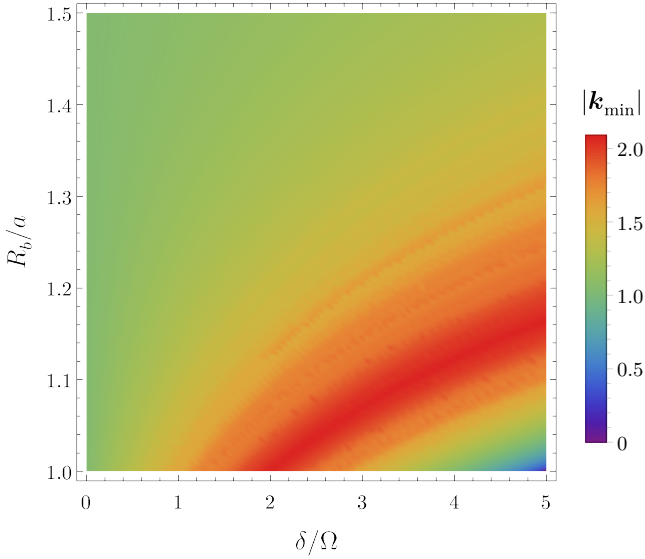

Figure S5. Absolute value of the wavevector corresponding to the dispersion minima of the $J_1$-$J_2$ Ising model [Eq. (S42)] as a function of $\delta/\Omega$ and $R_b/a$. While a smoothly varying $\boldsymbol{k}_{\min}$ seemingly suggests that vison condensation would lead to incommensurate ordering, this picture changes upon inclusion of a finite $J_3$, resulting in a nematic ordered phase.

where $\alpha = A, B$ is an index labeling the two sublattices of the honeycomb lattice. Introducing $\Psi_{\boldsymbol{k}} = (\mu^z_{\boldsymbol{k},A}, \mu^z_{\boldsymbol{k},B})^{\mathrm{T}}$, Eq. (S37) can be compactly written as

$$\tilde{\mathcal{H}}_0^{\mathrm{even}}(\boldsymbol{k}) = \Psi_{\boldsymbol{k}}^\dagger \, h(\boldsymbol{k}) \, \Psi_{\boldsymbol{k}}, \quad \text{with}$$

$$h(\boldsymbol{k}) = \begin{bmatrix} \dfrac{V_1}{8}\gamma(\boldsymbol{k}) & \left(\dfrac{\delta}{2}-V_1\right)\zeta(\boldsymbol{k}) \\ \left(\dfrac{\delta}{2}-V_1\right)\zeta(\boldsymbol{k})^* & \dfrac{V_1}{8}\gamma(\boldsymbol{k}) \end{bmatrix} \quad \text{(S42)}$$

where $\gamma(\boldsymbol{k})$ is defined in Eq. (S11) and

$$\zeta(\boldsymbol{k}) \equiv \frac{1}{2} + e^{i\sqrt{3}k_y}\cos(k_x). \quad \text{(S43)}$$

Diagonalizing Eq. (S42) yields the band structure for the visons. We find that, for the lower of the two bands, the location of the dispersion minima in the Brillouin zone changes smoothly as the parameters $(\delta/\Omega, R_b/a)$ are varied, as plotted in Fig. S5—however, this is likely an artifact of our simplistic model (S40). In fact, retaining the second-nearest-neighbor ($V_2$) interactions in the six-qubit term of Eq. (S37) would—at the mean-field level—lead to an Ising model with third-nearest-neighbor couplings, $J_3$, and renormalized values of $J_1$ and $J_2$. Roychowdhury et al. [7] showed that in such a model with a nonzero $J_3$ on top of $J_1$ and $J_2$, there is an extended region in parameter space where the minima of the spectrum occur at the three inequivalent M points defined by $\mathrm{M}_i = \boldsymbol{b}_i/2$. Therefore, pairwise condensation of the visons can result in a solid phase with ordering wavevectors at $\boldsymbol{b}_i$, which is nothing but the nematic phase of Rydberg atoms [10].

### 2. Odd $\mathbb{Z}_2$ spin liquid

The condition $\prod_{\ell' \in \bigcirc_i} \eta^z_{\ell'} = -1 \; \forall \; i$ in Eq. (S36) constrains any choice of a gauge for the odd QSL. However, in a suitably chosen gauge, we can write the Hamiltonian as

$$\tilde{\mathcal{H}}_0^{\mathrm{odd}} = J_1 \sum_{\langle i',j' \rangle} M_{i'j'}\mu^z_{i'}\mu^z_{j'} + J_2 \sum_{\langle\langle i',j' \rangle\rangle} M_{i'j'}\mu^z_{i'}\mu^z_{j'}, \quad \text{(S44)}$$

where $M_{i'j'} = -1$ if there exists a link $\ell'$ on the shortest path along the edges of the honeycomb lattice connecting $i'$ and $j'$ such that $\eta_{\ell'} = -1$; otherwise, $M_{i'j'} = +1$. The final theory for the visons thus resembles the fully frustrated Ising magnet on the honeycomb lattice [11] but with added interactions beyond nearest neighbors. One possible gauge choice is sketched in Fig. S6(a); since the $\eta_{\ell'}$ are now nonuniform, our unit cell has to be expanded to four sites. In units where $a=1$, the new lattice vectors are $\boldsymbol{d}_1 = (4,0)$, $\boldsymbol{d}_2 = (1,\sqrt{3})$, and we define $\boldsymbol{d}_3 = -\boldsymbol{d}_1 + \boldsymbol{d}_2$. The corresponding reciprocal lattice vectors are given by $\boldsymbol{g}_1 = \pi\left(1/2, -1/(2\sqrt{3})\right)$ and $\boldsymbol{g}_2 = \pi(0, 2/\sqrt{3})$. Using the shorthand $\xi_j = \exp\left(i\boldsymbol{k}\cdot\boldsymbol{d}_j\right)$, Eq. (S44) can be expressed as

$$\tilde{\mathcal{H}}_0^{\mathrm{odd}}(\boldsymbol{k}) = \Psi_{\boldsymbol{k}}^\dagger \, \tilde{h}(\boldsymbol{k}) \, \Psi_{\boldsymbol{k}}; \quad \Psi_{\boldsymbol{k}} = (\mu^z_{\boldsymbol{k},A}, \mu^z_{\boldsymbol{k},B}, \mu^z_{\boldsymbol{k},C}, \mu^z_{\boldsymbol{k},D})^{\mathrm{T}},$$

where

$$\begin{aligned}
\tilde{h}(\boldsymbol{k}) = &\frac{1}{2}\left(\frac{\delta}{2}-V_1\right)\begin{bmatrix} 0 & 1+\xi_2^* & 0 & \xi_1^* \\ 1+\xi_2 & 0 & 1 & 0 \\ 0 & 1 & 0 & 1-\xi_2^* \\ \xi_1 & 0 & 1-\xi_2 & 0 \end{bmatrix} \\
&+ \frac{V_1}{8}\begin{bmatrix} \xi_2+\xi_2^* & 0 & 1+\xi_1^*+\xi_2^*-\xi_3 & 0 \\ 0 & \xi_2+\xi_2^* & 0 & 1+\xi_1^*-\xi_2^*+\xi_3 \\ 1+\xi_1+\xi_2-\xi_3^* & 0 & -\xi_2-\xi_2^* & 0 \\ 0 & 1+\xi_1-\xi_2+\xi_3^* & 0 & -\xi_2-\xi_2^* \end{bmatrix}.
\end{aligned} \quad \text{(S45)}$$

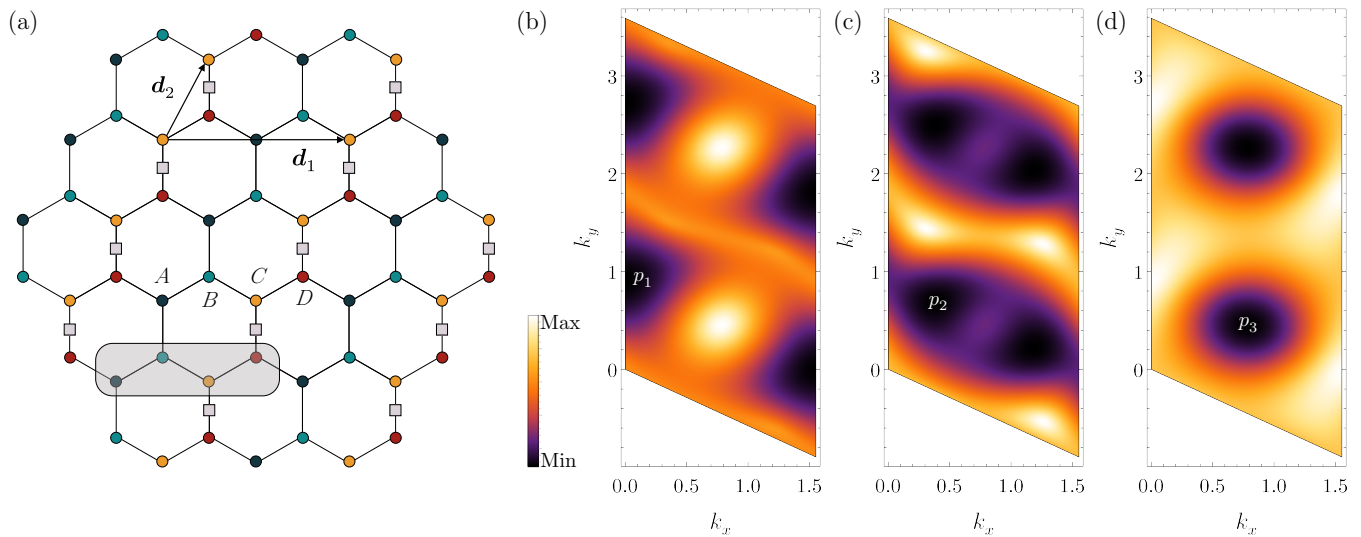

Figure S6. (a) The gauge chosen in writing the odd $\mathbb{Z}_2$ QSL Hamiltonian in Eq. (S45): here, $\eta_{\ell'} = -1$ on all links marked by the gray squares and equals $+1$ everywhere else. The resultant unit cell is composed of four sites (labeled by $A, B, C, D$), as indicated by the shaded box. (b–d) Dispersion of the lowest vison band for $\delta/\Omega = 4.0$, with (b) $R_b/a = 1.2$, (c) $R_b/a = 1.3$, and (d) $R_b/a = 1.6$, showing the three types of spectra and the locations of their minima in the Brillouin zone.

On diagonalizing this Hamiltonian, we find that the minima of the lowest vison band remain pinned to certain points in the Brillouin zone over extended regions of the $(\delta/\Omega, R_b/a)$-parameter space. In particular, we observe that for reasonable values of $\delta/\Omega$ and $R_b/a$, the dispersion can only be of one of three types—as illustrated in Fig. S6(b–d)—with their respective minima located at

I. $\boldsymbol{p}_1 \equiv \left(0, \frac{\pi}{2\sqrt{3}}\right)$, $\boldsymbol{p}_1 + \boldsymbol{g}_1$, $\boldsymbol{p}_1 + \frac{\boldsymbol{g}_2}{2}$, $\boldsymbol{p}_1 + \boldsymbol{g}_1 + \frac{\boldsymbol{g}_2}{2}$;

II. $\boldsymbol{p}_2 \equiv \left(\frac{\pi}{8}, \frac{\sqrt{3}\pi}{8}\right)$, $\boldsymbol{p}_2 + \frac{\boldsymbol{g}_1}{2}$, $\boldsymbol{p}_2 + \frac{\boldsymbol{g}_2}{2}$, $\boldsymbol{p}_2 + \frac{\boldsymbol{g}_1}{2} + \frac{\boldsymbol{g}_2}{2}$;

III. $\boldsymbol{p}_3 \equiv \left(\frac{\pi}{4}, \frac{\pi}{4\sqrt{3}}\right)$, $\boldsymbol{p}_3 + \frac{\boldsymbol{g}_2}{2}$. (S46)

As mentioned earlier, the momenta of these soft modes bear interesting consequences for the neighboring ordered phases. For instance, let us focus on the type-III spectrum of Fig. S6(d). Then, from Eq. (S46), we notice that

$$\left(\boldsymbol{p}_3 + \frac{\boldsymbol{g}_2}{2}\right) + \boldsymbol{p}_3 = \boldsymbol{b}_1 + (\boldsymbol{g}_2 - \boldsymbol{g}_1),$$
$$\left(\boldsymbol{p}_3 + \frac{\boldsymbol{g}_2}{2}\right) + \boldsymbol{p}_3 = \boldsymbol{b}_2 + (\boldsymbol{g}_1), \quad \text{(S47)}$$
$$\left(\boldsymbol{p}_3 + \frac{\boldsymbol{g}_2}{2}\right) + \boldsymbol{p}_3 = \boldsymbol{b}_3 + (-\boldsymbol{g}_1).$$

Since the right-hand sides of the equations above are simply the ordering peaks of the nematic solid (modulo the reciprocal lattice vectors indicated in brackets), the odd $\mathbb{Z}_2$ spin liquid can exhibit a transition into the nematic phase driven by the condensation of visons. Combinations of the form of Eq. (S47) that lead to nematic ordering can also be constructed from the momenta of the soft modes in the type-I spectrum.

Interestingly, we find that some other pairings of momenta, such as

$$\boldsymbol{p}_2 + \boldsymbol{p}_2 = \left(\frac{3\pi}{4}, \frac{\pi}{4\sqrt{3}}\right)(-\boldsymbol{g}_1),$$
$$\left(\boldsymbol{p}_3 + \frac{\boldsymbol{g}_2}{2}\right) + \boldsymbol{p}_3 = \left(\frac{\pi}{2}, \frac{\sqrt{3}\pi}{2}\right), \quad \text{(S48)}$$

reproduce a subset of the ordering wavevectors of a symmetry-breaking staggered phase [10]. However, it is believed that the transition between the odd $\mathbb{Z}_2$ QSL and the staggered solid is first-order, so a vison-condensation picture does not hold [8].

## SIV. FERMIONIC SPINON EXCITATIONS

The only quasiparticle excitation of the $\mathbb{Z}_2$ spin liquid that remains to be addressed is the fermionic spinon, which we now turn to in an Abrikosov fermion mean-field description. Starting with the familiar Rydberg Hamiltonian describing $N$ interacting atoms,

$$H_{\text{Ryd}} = \sum_{i=1}^{N} \left[ \frac{\Omega}{2} \left( |g\rangle_i \langle r| + |r\rangle_i \langle g| \right) - \delta |r\rangle_i \langle r| \right] + \frac{1}{2} \sum_{i,j} V_{i,j} |r\rangle_i \langle r| \otimes |r\rangle_j \langle r|, \tag{S49}$$

we rewrite the two-level system in terms of conventional spin-1/2 operators $\{S^x, S^y, S^z\}$ as

$$H = \sum_{i=1}^{N} \left[ \Omega S_i^z - \delta \left( S_i^x + \frac{1}{2} \right) \right] + \frac{1}{2} \sum_{i,j} V_{i,j} \left( S_i^x + \frac{1}{2} \right) \left( S_j^x + \frac{1}{2} \right). \tag{S50}$$

The atoms—or equivalently, the spins $S^a$—reside on the *sites* of the kagome lattice. Thus, despite our use of the indices $i, j$ to refer to the degrees of freedom in this section, we will be working on the $\ell$-lattice of our earlier discussion.

### A. Formalism

In the Abrikosov fermion representation [12–14], each spin is decomposed into fermions as

$$S_i^a = \frac{1}{2} f_{i,\alpha}^\dagger \rho_{\alpha\beta}^a f_{i,\beta}, \quad \sum_\alpha f_{i,\alpha}^\dagger f_{i,\alpha} = 1, \quad f_{i,\alpha} f_{i,\beta} \epsilon_{\alpha\beta} = 0, \tag{S51}$$

where $\{\rho^a\}$ are the usual Pauli matrices. In terms of the Abrikosov fermions, we obtain the following Hamiltonian:

$$H = H_0 + H_2 + H_4 + H_\lambda, \quad \text{with} \tag{S52}$$

$$H_0 = -\frac{\delta N}{2} + \sum_{i,j} \frac{V_{i,j}}{8}, \tag{S53}$$

$$H_2 = \sum_{i=1}^{N} \left[ \frac{\Omega}{2} f_{i,\alpha}^\dagger \rho_{\alpha\beta}^z f_{i,\beta} - \frac{\delta}{2} f_{i,\alpha}^\dagger \rho_{\alpha\beta}^x f_{i,\beta} \right] + \sum_{i,j} \frac{V_{i,j}}{8} \left[ f_{i,\alpha}^\dagger \rho_{\alpha\beta}^x f_{i,\beta} + f_{j,\alpha}^\dagger \rho_{\alpha\beta}^x f_{j,\beta} \right], \tag{S54}$$

$$H_4 = -\sum_{i,j} \frac{V_{i,j}}{8} (1 - \delta_{\alpha\beta}) \left[ f_{i,\alpha}^\dagger f_{j,\alpha}^\dagger f_{i,\beta} f_{j,\beta} + f_{i,\alpha}^\dagger f_{j,\beta}^\dagger f_{i,\beta} f_{j,\alpha} \right], \tag{S55}$$

$$H_\lambda = \sum_i \left[ \lambda_1 \left( f_{i,\alpha}^\dagger f_{i,\alpha} - 1 \right) + \left( \lambda_2 f_{i,\alpha} f_{i,\beta} \epsilon_{\alpha,\beta} + \text{H.c.} \right) \right], \tag{S56}$$

where $\lambda_1$ and $\lambda_2$ are Lagrange multipliers to impose the constraint introduced by the last two equations in (S51). Next, we decouple the quartic term to obtain a mean-field theory; this can be done as follows:

$$\begin{aligned}
f_{i,\alpha}^\dagger f_{j,\alpha}^\dagger f_{i,\beta} f_{j,\beta} &= f_{i,\alpha}^\dagger f_{j,\alpha}^\dagger \left\langle f_{i,\beta} f_{j,\beta} \right\rangle + \left\langle f_{i,\alpha}^\dagger f_{j,\alpha}^\dagger \right\rangle f_{i,\beta} f_{j,\beta} - \left\langle f_{i,\alpha}^\dagger f_{j,\alpha}^\dagger \right\rangle \left\langle f_{i,\beta} f_{j,\beta} \right\rangle \\
&+ f_{i,\alpha}^\dagger f_{j,\beta} \left\langle f_{j,\alpha}^\dagger f_{i,\beta} \right\rangle + \left\langle f_{i,\alpha}^\dagger f_{j,\beta} \right\rangle f_{j,\alpha}^\dagger f_{i,\beta} - \left\langle f_{i,\alpha}^\dagger f_{j,\beta} \right\rangle \left\langle f_{j,\alpha}^\dagger f_{i,\beta} \right\rangle \\
&- f_{i,\alpha}^\dagger f_{i,\beta} \left\langle f_{j,\alpha}^\dagger f_{j,\beta} \right\rangle - \left\langle f_{i,\alpha}^\dagger f_{i,\beta} \right\rangle f_{j,\alpha}^\dagger f_{j,\beta} + \left\langle f_{i,\alpha}^\dagger f_{i,\beta} \right\rangle \left\langle f_{j,\alpha}^\dagger f_{j,\beta} \right\rangle,
\end{aligned} \tag{S57}$$

$$\begin{aligned}
f_{i,\alpha}^\dagger f_{j,\beta}^\dagger f_{i,\beta} f_{j,\alpha} &= f_{i,\alpha}^\dagger f_{j,\beta}^\dagger \left\langle f_{i,\beta} f_{j,\alpha} \right\rangle + \left\langle f_{i,\alpha}^\dagger f_{j,\beta}^\dagger \right\rangle f_{i,\beta} f_{j,\alpha} - \left\langle f_{i,\alpha}^\dagger f_{j,\beta}^\dagger \right\rangle \left\langle f_{i,\beta} f_{j,\alpha} \right\rangle \\
&+ f_{i,\alpha}^\dagger f_{j,\alpha} \left\langle f_{j,\beta}^\dagger f_{i,\beta} \right\rangle + \left\langle f_{i,\alpha}^\dagger f_{j,\alpha} \right\rangle f_{j,\beta}^\dagger f_{i,\beta} - \left\langle f_{i,\alpha}^\dagger f_{j,\alpha} \right\rangle \left\langle f_{j,\beta}^\dagger f_{i,\beta} \right\rangle \\
&- f_{i,\alpha}^\dagger f_{i,\beta} \left\langle f_{j,\beta}^\dagger f_{j,\alpha} \right\rangle - \left\langle f_{i,\alpha}^\dagger f_{i,\beta} \right\rangle f_{j,\beta}^\dagger f_{j,\alpha} + \left\langle f_{i,\alpha}^\dagger f_{i,\beta} \right\rangle \left\langle f_{j,\beta}^\dagger f_{j,\alpha} \right\rangle.
\end{aligned} \tag{S58}$$

For simplicity, let us introduce the notation

$$\Delta_{ij}^{\alpha\beta} = \left\langle f_{i,\alpha} f_{j,\beta} \right\rangle, \quad t_{ij}^{\alpha\beta} = \left\langle f_{i,\alpha}^\dagger f_{j,\beta} \right\rangle, \tag{S59}$$

to represent the pairing and hopping amplitudes, respectively, in terms of which,

$$\begin{aligned}
f_{i,\alpha}^\dagger f_{j,\alpha}^\dagger f_{i,\beta} f_{j,\beta} &= -t_{ii}^{\alpha\beta} f_{j,\alpha}^\dagger f_{j,\beta} - t_{jj}^{\alpha\beta} f_{i,\alpha}^\dagger f_{i,\beta} + \left( t_{ji}^{\alpha\beta} f_{i,\alpha}^\dagger f_{j,\beta} + \Delta_{ij}^{\beta\beta} f_{i,\alpha}^\dagger f_{j,\alpha}^\dagger + \text{H.c.} \right) \\
&+ \Delta_{ij}^{\beta\beta} (\Delta_{ij}^{\alpha\alpha})^* - t_{ij}^{\alpha\beta} t_{ji}^{\alpha\beta} + t_{ii}^{\alpha\beta} t_{jj}^{\alpha\beta},
\end{aligned} \tag{S60}$$

$$f_{i,\alpha}^\dagger f_{j,\beta}^\dagger f_{i,\beta} f_{j,\alpha} = -t_{ii}^{\beta\alpha} f_{j,\alpha}^\dagger f_{j,\beta} - t_{jj}^{\beta\alpha} f_{i,\alpha}^\dagger f_{i,\beta} + \left( t_{ji}^{\beta\beta} f_{i,\alpha}^\dagger f_{j,\alpha} + \Delta_{ij}^{\beta\alpha} f_{i,\alpha}^\dagger f_{j,\beta}^\dagger + \text{H.c.} \right)$$
$$+ \Delta_{ij}^{\beta\alpha}(\Delta_{ij}^{\alpha\beta})^* - t_{ij}^{\alpha\alpha} t_{ji}^{\beta\beta} + t_{ii}^{\alpha\beta} t_{jj}^{\beta\alpha}. \tag{S61}$$

Consequently, Eq. (S55) then reads

$$H_4 = \bar{H}_0 + \bar{H}_2, \quad \text{with} \tag{S62}$$

$$\bar{H}_0 = -\sum_{i,j} \frac{V_{i,j}}{8} \left(1 - \delta_{\alpha\beta}\right) \left[ \Delta_{ij}^{\beta\beta}(\Delta_{ij}^{\alpha\alpha})^* - t_{ij}^{\alpha\beta} t_{ji}^{\alpha\beta} + t_{ii}^{\alpha\beta} t_{jj}^{\alpha\beta} + \Delta_{ij}^{\beta\alpha}(\Delta_{ij}^{\alpha\beta})^* - t_{ij}^{\alpha\alpha} t_{ji}^{\beta\beta} + t_{ii}^{\alpha\beta} t_{jj}^{\beta\alpha} \right], \tag{S63}$$

$$\bar{H}_2 = -\sum_{i,j} \frac{V_{i,j}}{8} \left(1 - \delta_{\alpha\beta}\right) \left[ -t_{ii}^{\alpha\beta} f_{j,\alpha}^\dagger f_{j,\beta} - t_{jj}^{\alpha\beta} f_{i,\alpha}^\dagger f_{i,\beta} + \left( t_{ji}^{\alpha\beta} f_{i,\alpha}^\dagger f_{j,\beta} + \Delta_{ij}^{\beta\beta} f_{i,\alpha}^\dagger f_{j,\alpha}^\dagger + \text{H.c.} \right) \right.$$
$$\left. - t_{ii}^{\beta\alpha} f_{j,\alpha}^\dagger f_{j,\beta} - t_{jj}^{\beta\alpha} f_{i,\alpha}^\dagger f_{i,\beta} + \left( t_{ji}^{\beta\beta} f_{i,\alpha}^\dagger f_{j,\alpha} + \Delta_{ij}^{\beta\alpha} f_{i,\alpha}^\dagger f_{j,\beta}^\dagger + \text{H.c.} \right) \right], \tag{S64}$$

$$H = H_0 + \bar{H}_0 + H_2 + \bar{H}_2 + H_\lambda. \tag{S65}$$

Now, we only have a bilinear Hamiltonian to deal with in Eq. (S65). Before proceeding further, however, we have to specify the action of lattice translation symmetries on the expectation values $t_{ij}^{\alpha\beta}$ and $\Delta_{ij}^{\alpha\beta}$—this defines a mean-field *ansatz*. Here, we consider a simple ansatz with only nearest-neighbor mean-field bonds, chosen such that there is no flux piercing either the triangular or the honeycomb plaquettes of the kagome lattice [15, 16].

At this stage, it is convenient to transform our Hamiltonian to Fourier space using

$$f_{i,\alpha} = \frac{1}{\sqrt{N}} \sum_{\boldsymbol{k}} f_{\boldsymbol{k},\alpha} e^{-i\boldsymbol{k}\cdot\boldsymbol{r}_i}. \tag{S66}$$

Note that the kagome lattice is generated by $\boldsymbol{R}_1 = (2a, 0)$ and $\boldsymbol{R}_2 = (a, \sqrt{3}a)$, where $a$ is the lattice spacing, so in momentum space, the reciprocal lattice vectors are $\boldsymbol{b}_1 = (1, -1/\sqrt{3})(\pi/a)$ and $\boldsymbol{b}_2 = (0, 2/\sqrt{3})(\pi/a)$. The bilinear Hamiltonian, $H_2 + \bar{H}_2 + H_\lambda$, can be expressed as

$$H_{2\boldsymbol{k}} = \frac{1}{2} \sum_{\boldsymbol{k}} \Psi_{\boldsymbol{k}}^\dagger \mathcal{M}_{\boldsymbol{k}} \Psi_{\boldsymbol{k}}, \tag{S67}$$

where $\Psi_{\boldsymbol{k}}^\dagger = \left( f_{\boldsymbol{k}A,\uparrow}^\dagger, f_{\boldsymbol{k}A,\downarrow}^\dagger, f_{\boldsymbol{k}B,\uparrow}^\dagger, f_{\boldsymbol{k}B,\downarrow}^\dagger, f_{\boldsymbol{k}C,\uparrow}^\dagger, f_{\boldsymbol{k}C,\downarrow}^\dagger, f_{-\boldsymbol{k}A,\uparrow}, f_{-\boldsymbol{k}A,\downarrow}, f_{-\boldsymbol{k}B,\uparrow}, f_{-\boldsymbol{k}B,\downarrow}, f_{-\boldsymbol{k}C,\uparrow}, f_{-\boldsymbol{k}C,\downarrow} \right)$, with $A, B$, and $C$ being the labels for the three sublattices. Here, $\mathcal{M}_{\boldsymbol{k}}$ is a 12×12 matrix formed using two 6×6 matrices, $T_{\boldsymbol{k}}$ and $\Delta_{\boldsymbol{k}}$, which contain the hopping and pairing parts, respectively. The $T_{\boldsymbol{k}}$ and $\Delta_{\boldsymbol{k}}$ matrices are in turn composed of 2×2 matrices each:

$$\mathcal{M}_{\boldsymbol{k}} = \begin{bmatrix} T_{\boldsymbol{k}} & \Delta_{\boldsymbol{k}} \\ \Delta_{\boldsymbol{k}}^\dagger & -T_{-\boldsymbol{k}}^{\mathrm{T}} \end{bmatrix}, \quad \text{with} \quad T_{\boldsymbol{k}} = \begin{bmatrix} T_{AA} & T_{AB} & T_{AC} \\ T_{AB}^\dagger & T_{BB} & T_{BC} \\ T_{AC}^\dagger & T_{BC}^\dagger & T_{CC} \end{bmatrix}, \quad \Delta_{\boldsymbol{k}} = \begin{bmatrix} \Delta_{\boldsymbol{k}}^{AA} & \Delta_{\boldsymbol{k}}^{AB} & \Delta_{\boldsymbol{k}}^{AC} \\ -\Delta_{-\boldsymbol{k}}^{AB} & \Delta_{\boldsymbol{k}}^{BB} & \Delta_{\boldsymbol{k}}^{BC} \\ -\Delta_{-\boldsymbol{k}}^{AC} & -\Delta_{-\boldsymbol{k}}^{BC} & \Delta_{\boldsymbol{k}}^{CC} \end{bmatrix}. \tag{S68}$$

The individual entries are

$$T_{AA} = T_{BB} = T_{CC} = \frac{\Omega}{2}\rho^z - \frac{\delta}{2}\rho^x + \frac{V_1 z_1 + V_2 z_2}{8}\rho^x + \lambda_1 \mathbb{1} + \frac{V_1 z_1 + V_2 z_2}{8}(t_0^{\uparrow\downarrow} + t_0^{\downarrow\uparrow})\rho^x, \tag{S69}$$

$$T_{AB} = -\frac{V_1}{8}(2\cos k_2) \begin{bmatrix} t_1^{\downarrow\downarrow} & t_1^{\uparrow\downarrow} \\ t_1^{\downarrow\uparrow} & t_1^{\uparrow\uparrow} \end{bmatrix} - \frac{V_2}{8}(2\cos k_2') \begin{bmatrix} t_2^{\downarrow\downarrow *} & t_2^{\downarrow\uparrow *} \\ t_2^{\uparrow\downarrow *} & t_2^{\uparrow\uparrow *} \end{bmatrix}, \tag{S70}$$

$$T_{AC} = -\frac{V_1}{8}(2\cos k_1) \begin{bmatrix} t_1^{\downarrow\downarrow *} & t_1^{\downarrow\uparrow *} \\ t_1^{\uparrow\downarrow *} & t_1^{\uparrow\uparrow *} \end{bmatrix} - \frac{V_2}{8}(2\cos k_1') \begin{bmatrix} t_2^{\downarrow\downarrow} & t_2^{\uparrow\downarrow} \\ t_2^{\downarrow\uparrow} & t_2^{\uparrow\uparrow} \end{bmatrix}, \tag{S71}$$

$$T_{BC} = -\frac{V_1}{8}(2\cos k_3) \begin{bmatrix} t_1^{\downarrow\downarrow} & t_1^{\uparrow\downarrow} \\ t_1^{\downarrow\uparrow} & t_1^{\uparrow\uparrow} \end{bmatrix} - \frac{V_2}{8}(2\cos k_3') \begin{bmatrix} t_2^{\downarrow\downarrow *} & t_2^{\downarrow\uparrow *} \\ t_2^{\uparrow\downarrow *} & t_2^{\uparrow\uparrow *} \end{bmatrix}, \tag{S72}$$

$$\Delta_{\boldsymbol{k}}^{AA} = \Delta_{\boldsymbol{k}}^{BB} = \Delta_{\boldsymbol{k}}^{CC} = 2i\,\lambda_2^*\,\rho^y, \tag{S73}$$

$$\Delta_{\boldsymbol{k}}^{AB} = -\frac{V_1}{8}(2\cos k_2) \begin{bmatrix} -\Delta_1^{\downarrow\downarrow} & -\Delta_1^{\uparrow\downarrow} \\ -\Delta_1^{\downarrow\uparrow} & -\Delta_1^{\uparrow\uparrow} \end{bmatrix} - \frac{V_2}{8}(2\cos k_2') \begin{bmatrix} \Delta_2^{\downarrow\downarrow} & \Delta_2^{\downarrow\uparrow} \\ \Delta_2^{\uparrow\downarrow} & \Delta_2^{\uparrow\uparrow} \end{bmatrix}, \tag{S74}$$

$$\Delta_{\boldsymbol{k}}^{AC} = -\frac{V_1}{8}(2\cos k_1) \begin{bmatrix} \Delta_1^{\downarrow\downarrow} & \Delta_1^{\downarrow\uparrow} \\ \Delta_1^{\uparrow\downarrow} & \Delta_1^{\uparrow\uparrow} \end{bmatrix} - \frac{V_2}{8}(2\cos k_1') \begin{bmatrix} -\Delta_2^{\downarrow\downarrow} & -\Delta_2^{\downarrow\uparrow} \\ -\Delta_2^{\uparrow\downarrow} & -\Delta_2^{\uparrow\uparrow} \end{bmatrix}, \tag{S75}$$

$$\Delta_{\boldsymbol{k}}^{BC} = -\frac{V_1}{8}(2\cos k_3) \begin{bmatrix} -\Delta_1^{\downarrow\downarrow} & -\Delta_1^{\uparrow\downarrow} \\ -\Delta_1^{\downarrow\uparrow} & -\Delta_1^{\uparrow\uparrow} \end{bmatrix} - \frac{V_2}{8}(2\cos k_3') \begin{bmatrix} \Delta_2^{\downarrow\downarrow} & \Delta_2^{\downarrow\uparrow} \\ \Delta_2^{\uparrow\downarrow} & \Delta_2^{\uparrow\uparrow} \end{bmatrix}. \tag{S76}$$

Let us briefly review the notation employed in the equations above. Here, $z_1$ and $z_2$ are the number of nearest and next-nearest neighbors, respectively. Additionally,

$$t_0^{\alpha\alpha} = \left\langle f_{A,\alpha}^\dagger f_{A,\alpha} \right\rangle = \left\langle f_{B,\alpha}^\dagger f_{B,\alpha} \right\rangle = \left\langle f_{C,\alpha}^\dagger f_{C,\alpha} \right\rangle, \tag{S77}$$

is the onsite mean-field parameter,

$$t_1^{\alpha\beta} = \left\langle f_{B,\alpha}^\dagger f_{A,\beta} \right\rangle = \left\langle f_{A,\alpha}^\dagger f_{C,\beta} \right\rangle = \left\langle f_{C,\alpha}^\dagger f_{B,\beta} \right\rangle, \tag{S78}$$

$$\Delta_1^{\alpha\beta} = \left\langle f_{B,\alpha} f_{A,\beta} \right\rangle = \left\langle f_{A,\alpha} f_{C,\beta} \right\rangle = \left\langle f_{C,\alpha} f_{B,\beta} \right\rangle, \tag{S79}$$

are the mean-field parameters defined on nearest-neighbor bonds, while

$$t_2^{\alpha\beta} = \left\langle f_{A,\alpha}^\dagger f_{B,\beta} \right\rangle = \left\langle f_{B,\alpha}^\dagger f_{C,\beta} \right\rangle = \left\langle f_{C,\alpha}^\dagger f_{A,\beta} \right\rangle, \tag{S80}$$

$$\Delta_2^{\alpha\beta} = \left\langle f_{A,\alpha} f_{B,\beta} \right\rangle = \left\langle f_{B,\alpha} f_{C,\beta} \right\rangle = \left\langle f_{C,\alpha} f_{A,\beta} \right\rangle, \tag{S81}$$

are defined on next-nearest-neighbor bonds. All these parameters are further constrained by the imposition of time-reversal symmetry. Lastly, we have also introduced the shorthand,

$$k_1 \equiv \boldsymbol{k} \cdot \boldsymbol{a}_1 = k_x, \quad k_2 \equiv \boldsymbol{k} \cdot \boldsymbol{a}_2 = \frac{k_x}{2} + \frac{\sqrt{3}k_y}{2}, \quad k_3 \equiv \boldsymbol{k} \cdot \boldsymbol{a}_3 = -\frac{k_x}{2} + \frac{\sqrt{3}k_y}{2},$$

$$k_1' \equiv \boldsymbol{k} \cdot \boldsymbol{a}_1' = \sqrt{3}k_y, \quad k_2' \equiv \boldsymbol{k} \cdot \boldsymbol{a}_2' = -\frac{3k_x}{2} + \frac{\sqrt{3}k_y}{2}, \quad k_3' \equiv \boldsymbol{k} \cdot \boldsymbol{a}_3' = \frac{3k_x}{2} + \frac{\sqrt{3}k_y}{2}. \tag{S82}$$

The vectors $\boldsymbol{a}_1$, $\boldsymbol{a}_2$ and $\boldsymbol{a}_3$ connect nearest neighbors $A$–$C$, $A$–$B$ and $C$–$B$, respectively, while, $\boldsymbol{a}_1'$, $\boldsymbol{a}_2'$ and $\boldsymbol{a}_3'$ connect the next-nearest neighbors $A$–$C$, $A$–$B$ and $C$–$B$, respectively.

With these definitions aside, we now briefly describe the procedure to diagonalize the matrix $\mathcal{M}_{\boldsymbol{k}}$ in Eq. (S67) and obtain a self-consistent solution for a quantum spin liquid state. To start, we initialize the mean-field parameters $t$ and $\Delta$ with random values. The actual diagonalization is implemented via a unitary transformation:

$$\Psi_{\boldsymbol{k}} = U_p \Phi_{\boldsymbol{k}}, \quad U_p = \begin{bmatrix} U & V \\ V^* & U^* \end{bmatrix}, \quad \Omega_{\boldsymbol{k}} = U_p^\dagger \mathcal{M}_{\boldsymbol{k}} U_p, \tag{S83}$$

where $\Phi_{\boldsymbol{k}}$ encapsulates the Bogoliubov operators as $\Phi_{\boldsymbol{k}}^\dagger = \left(\tau_{\boldsymbol{k}1}^\dagger \tau_{\boldsymbol{k}2}^\dagger \tau_{\boldsymbol{k}3}^\dagger \tau_{\boldsymbol{k}4}^\dagger \tau_{\boldsymbol{k}5}^\dagger \tau_{\boldsymbol{k}6}^\dagger \tau_{-\boldsymbol{k}1} \tau_{-\boldsymbol{k}2} \tau_{-\boldsymbol{k}3} \tau_{-\boldsymbol{k}4} \tau_{-\boldsymbol{k}5} \tau_{-\boldsymbol{k}6}\right)$, $U_p$ is a $12\times 12$ unitary matrix consisting of two $6\times 6$ matrices ($U$ and $V$), and $\Omega_{\boldsymbol{k}}$ is the diagonal matrix composed of the eigenvalues of $\mathcal{M}_{\boldsymbol{k}}$. As is easily seen from the definition (S83), the matrix $U_p$ contains the eigenvectors of $\mathcal{M}_{\boldsymbol{k}}$ as column vectors. Using the relations above, we can calculate the mean-field parameters $\langle f^\dagger f \rangle$ and $\langle ff \rangle$. We then iterate this procedure till we achieve convergence for the mean-field parameter values. Note that in this process, we also have to optimize for the self-consistent values of the chemical potentials such that the constraint (S51) is satisfied.

The key step in going from one iteration to the next is evaluating the different expectation values $\langle f^\dagger f \rangle$ and $\langle ff \rangle$. In practice, using the explicit expressions for the $f$ fermions in terms of the $\tau$ operators,

$$f_{\boldsymbol{k}A,\uparrow} = \sum_{i=1}^{6} \left[ u_{1i}(\boldsymbol{k}) \tau_{\boldsymbol{k},i} + v_{1i}(\boldsymbol{k}) \tau_{-\boldsymbol{k},i}^\dagger \right], \quad f_{\boldsymbol{k}A,\downarrow} = \sum_{i=1}^{6} \left[ u_{2i}(\boldsymbol{k}) \tau_{\boldsymbol{k},i} + v_{2i}(\boldsymbol{k}) \tau_{-\boldsymbol{k},i}^\dagger \right],$$

$$f_{\boldsymbol{k}B,\uparrow} = \sum_{i=1}^{6} \left[ u_{3i}(\boldsymbol{k})\tau_{\boldsymbol{k},i} + v_{3i}(\boldsymbol{k})\tau^\dagger_{-\boldsymbol{k},i} \right], \quad f_{\boldsymbol{k}B,\downarrow} = \sum_{i=1}^{6} \left[ u_{4i}(\boldsymbol{k})\tau_{\boldsymbol{k},i} + v_{4i}(\boldsymbol{k})\tau^\dagger_{-\boldsymbol{k},i} \right],$$

$$f_{\boldsymbol{k}C,\uparrow} = \sum_{i=1}^{6} \left[ u_{5i}(\boldsymbol{k})\tau_{\boldsymbol{k},i} + v_{5i}(\boldsymbol{k})\tau^\dagger_{-\boldsymbol{k},i} \right], \quad f_{\boldsymbol{k}C,\downarrow} = \sum_{i=1}^{6} \left[ u_{6i}(\boldsymbol{k})\tau_{\boldsymbol{k},i} + v_{6i}(\boldsymbol{k})\tau^\dagger_{-\boldsymbol{k},i} \right],$$

we can efficiently compute all the mean-field parameters in terms of the Bogoliubov coefficients as,

$$t_0^{\uparrow\uparrow} = \frac{1}{N} \sum_{\boldsymbol{k}} \left( \sum_{i=1}^{6} |v_{1i}(\boldsymbol{k})|^2 \right) = \frac{1}{N} \sum_{\boldsymbol{k}} \left( \sum_{i=1}^{6} |v_{3i}(\boldsymbol{k})|^2 \right) = \frac{1}{N} \sum_{\boldsymbol{k}} \left( \sum_{i=1}^{6} |v_{5i}(\boldsymbol{k})|^2 \right),$$

$$t_0^{\downarrow\downarrow} = \frac{1}{N} \sum_{\boldsymbol{k}} \left( \sum_{i=1}^{6} |v_{2i}(\boldsymbol{k})|^2 \right) = \frac{1}{N} \sum_{\boldsymbol{k}} \left( \sum_{i=1}^{6} |v_{4i}(\boldsymbol{k})|^2 \right) = \frac{1}{N} \sum_{\boldsymbol{k}} \left( \sum_{i=1}^{6} |v_{6i}(\boldsymbol{k})|^2 \right),$$

$$t_1^{\uparrow\uparrow} = \frac{1}{N} \sum_{\boldsymbol{k}} \cos k_2 \left( \sum_{i=1}^{6} v_{3i}^*(\boldsymbol{k}) v_{1i}(k) \right) = \frac{1}{N} \sum_{\boldsymbol{k}} \cos k_1 \left( \sum_{i=1}^{6} v_{1i}^*(\boldsymbol{k}) v_{5i}(k) \right)$$

$$= \frac{1}{N} \sum_{\boldsymbol{k}} \cos k_3 \left( \sum_{i=1}^{6} v_{5i}^*(\boldsymbol{k}) v_{3i}(k) \right),$$

$$t_1^{\downarrow\downarrow} = \frac{1}{N} \sum_{\boldsymbol{k}} \cos k_2 \left( \sum_{i=1}^{6} v_{4i}^*(\boldsymbol{k}) v_{2i}(k) \right) = \frac{1}{N} \sum_{\boldsymbol{k}} \cos k_1 \left( \sum_{i=1}^{6} v_{2i}^*(\boldsymbol{k}) v_{6i}(k) \right)$$

$$= \frac{1}{N} \sum_{\boldsymbol{k}} \cos k_3 \left( \sum_{i=1}^{6} v_{6i}^*(\boldsymbol{k}) v_{4i}(k) \right),$$

$$t_1^{\uparrow\downarrow} = \frac{1}{N} \sum_{\boldsymbol{k}} \cos k_2 \left( \sum_{i=1}^{6} v_{3i}^*(\boldsymbol{k}) v_{2i}(k) \right) = \frac{1}{N} \sum_{\boldsymbol{k}} \cos k_1 \left( \sum_{i=1}^{6} v_{1i}^*(\boldsymbol{k}) v_{6i}(k) \right)$$

$$= \frac{1}{N} \sum_{\boldsymbol{k}} \cos k_3 \left( \sum_{i=1}^{6} v_{5i}^*(\boldsymbol{k}) v_{4i}(k) \right),$$

$$t_1^{\downarrow\uparrow} = \frac{1}{N} \sum_{\boldsymbol{k}} \cos k_2 \left( \sum_{i=1}^{6} v_{4i}^*(\boldsymbol{k}) v_{1i}(k) \right) = \frac{1}{N} \sum_{\boldsymbol{k}} \cos k_1 \left( \sum_{i=1}^{6} v_{2i}^*(\boldsymbol{k}) v_{5i}(k) \right)$$

$$= \frac{1}{N} \sum_{\boldsymbol{k}} \cos k_3 \left( \sum_{i=1}^{6} v_{6i}^*(\boldsymbol{k}) v_{3i}(k) \right),$$

$$\Delta_1^{\uparrow\uparrow} = \frac{1}{N} \sum_{\boldsymbol{k}} \cos k_2 \left( \sum_{i=1}^{6} u_{3i}(\boldsymbol{k}) v_{1i}(-\boldsymbol{k}) \right) = \frac{1}{N} \sum_{\boldsymbol{k}} \cos k_1 \left( \sum_{i=1}^{6} u_{1i}(\boldsymbol{k}) v_{5i}(-\boldsymbol{k}) \right)$$

$$= \frac{1}{N} \sum_{\boldsymbol{k}} \cos k_3 \left( \sum_{i=1}^{6} u_{5i}(\boldsymbol{k}) v_{3i}(-\boldsymbol{k}) \right),$$

$$\Delta_1^{\downarrow\downarrow} = \frac{1}{N} \sum_{\boldsymbol{k}} \cos k_2 \left( \sum_{i=1}^{6} u_{4i}(\boldsymbol{k}) v_{2i}(-\boldsymbol{k}) \right) = \frac{1}{N} \sum_{\boldsymbol{k}} \cos k_1 \left( \sum_{i=1}^{6} u_{2i}(\boldsymbol{k}) v_{6i}(-\boldsymbol{k}) \right)$$

$$= \frac{1}{N} \sum_{\boldsymbol{k}} \cos k_3 \left( \sum_{i=1}^{6} u_{6i}(\boldsymbol{k}) v_{4i}(-\boldsymbol{k}) \right),$$

$$\Delta_1^{\uparrow\downarrow} = \frac{1}{N} \sum_{\boldsymbol{k}} \cos k_2 \left( \sum_{i=1}^{6} u_{3i}(\boldsymbol{k}) v_{2i}(-\boldsymbol{k}) \right) = \frac{1}{N} \sum_{\boldsymbol{k}} \cos k_1 \left( \sum_{i=1}^{6} u_{1i}(\boldsymbol{k}) v_{6i}(-\boldsymbol{k}) \right)$$

$$= \frac{1}{N} \sum_{\boldsymbol{k}} \cos k_3 \left( \sum_{i=1}^{6} u_{5i}(\boldsymbol{k}) v_{4i}(-\boldsymbol{k}) \right),$$

$$\Delta_1^{\downarrow\uparrow} = \frac{1}{N}\sum_{\boldsymbol{k}} \cos k_2 \left(\sum_{i=1}^{6} u_{4i}(\boldsymbol{k})v_{1i}(-\boldsymbol{k})\right) = \frac{1}{N}\sum_{\boldsymbol{k}} \cos k_1 \left(\sum_{i=1}^{6} u_{2i}(\boldsymbol{k})v_{5i}(-\boldsymbol{k})\right)$$
$$= \frac{1}{N}\sum_{\boldsymbol{k}} \cos k_3 \left(\sum_{i=1}^{6} u_{6i}(\boldsymbol{k})v_{3i}(-\boldsymbol{k})\right).$$

As an example, we tabulate here the mean-field parameters obtained self-consistently at $\delta/\Omega = 4.0$, $R_b/a = 1.6$, which were used to generate the fermionic spectrum in Fig. 4(a). We observe that the non-SU(2)-symmetric terms are generically small for hoppings but can be significant for pairings.

| Parameter | Real part | Imaginary part |
|---|---|---|
| $t_1^{\uparrow\uparrow}$ | $-0.1178$ | $0$ |
| $t_1^{\downarrow\downarrow}$ | $-0.1287$ | $0$ |
| $t_1^{\uparrow\downarrow}$ | $+3.5605 \times 10^{-5}$ | $0$ |
| $t_1^{\downarrow\uparrow}$ | $+3.6064 \times 10^{-5}$ | $0$ |
| $\Delta_1^{\uparrow\uparrow}$ | $-1.1263$ | $+9.0143 \times 10^{-2}$ |
| $\Delta_1^{\downarrow\downarrow}$ | $+1.1770$ | $-9.4206 \times 10^{-2}$ |
| $\Delta_1^{\uparrow\downarrow}$ | $-7.9412 \times 10^{-3}$ | $+6.3559 \times 10^{-4}$ |
| $\Delta_1^{\downarrow\uparrow}$ | $-7.9412 \times 10^{-3}$ | $+6.3559 \times 10^{-4}$ |

### B. Static structure factor

In order to determine an experimental signature of the fermionic quasiparticles, we now compute the static structure factor,

$$\mathcal{S}(\boldsymbol{q}) = \frac{1}{N}\sum_{ij} \left\langle \vec{S}_i \cdot \vec{S}_j \right\rangle e^{-i\boldsymbol{q}\cdot(\boldsymbol{r}_i-\boldsymbol{r}_j)} = \left\langle \vec{S}_{\boldsymbol{q}} \cdot \vec{S}_{-\boldsymbol{q}} \right\rangle. \quad \text{(S84)}$$

Recall that there are three sublattices on the kagome lattice, in terms of which,

$$\left\langle \vec{S}_{\boldsymbol{q}} \cdot \vec{S}_{-\boldsymbol{q}} \right\rangle = \sum_{l,m \in A,B,C} \left\langle \vec{S}_{\boldsymbol{q},l} \cdot \vec{S}_{-\boldsymbol{q},m} \right\rangle. \quad \text{(S85)}$$

Furthermore, using the relation

$$S_{\boldsymbol{q},l}^a = \frac{1}{\sqrt{N}}\sum_{\boldsymbol{k}_1} \frac{1}{2} f_{\boldsymbol{k}_1 l,\alpha}^\dagger \rho_{\alpha\beta}^a f_{(\boldsymbol{k}_1-\boldsymbol{q})l,\beta}, \quad \text{(S86)}$$

we have

$$\left\langle \vec{S}_{\boldsymbol{q},l} \cdot \vec{S}_{-\boldsymbol{q},m} \right\rangle = \frac{3}{4N}\sum_{\boldsymbol{k}_1} \left\langle f_{\boldsymbol{k}_1 l,\alpha}^\dagger f_{\boldsymbol{k}_1 m,\alpha} \right\rangle \delta_{l,m}$$
$$- \frac{1}{2N}\sum_{\boldsymbol{k}_1,\boldsymbol{k}_2} \left\langle f_{\boldsymbol{k}_1 l,\alpha}^\dagger f_{\boldsymbol{k}_2 m,\beta}^\dagger f_{(\boldsymbol{k}_1-\boldsymbol{q})l,\beta} f_{(\boldsymbol{k}_2+\boldsymbol{q})m,\alpha} \right\rangle$$
$$+ \frac{1}{4N}\sum_{\boldsymbol{k}_1,\boldsymbol{k}_2} \left\langle f_{\boldsymbol{k}_1 l,\alpha}^\dagger f_{\boldsymbol{k}_2 m,\beta}^\dagger f_{(\boldsymbol{k}_1-\boldsymbol{q})l,\alpha} f_{(\boldsymbol{k}_2+\boldsymbol{q})m,\beta} \right\rangle. \quad \text{(S87)}$$

At this stage, we evaluate the quartic terms in mean-field fashion using all possible mean-field decompositions, just as we did when solving the original Hamiltonian. Concisely, the expectation values listed above will depend on the Bogoliubov coefficients that were obtained earlier. Compiling these pieces, we obtain the static structure factor shown in Fig. 4(b). Note that $\mathcal{S}(\boldsymbol{q})$ exhibits sixfold rotational symmetry and does not have any sharp Bragg peaks (indicating the absence of long-range order)—this may be regarded as the hallmark of a quantum spin liquid.



\end{document}